\documentclass[fleqn,usenatbib]{mnras}

\usepackage{newtxtext,newtxmath}

\usepackage[T1]{fontenc}

\DeclareRobustCommand{\VAN}[3]{#2}
\let\VANthebibliography\thebibliography
\def\thebibliography{\DeclareRobustCommand{\VAN}[3]{##3}\VANthebibliography}

\usepackage{graphicx}	
\usepackage{amsmath}	
\usepackage{siunitx}
\DeclareSIUnit\angstrom{\text{Å}}

\defcitealias{Boselli2019}{B+19}
\defcitealias{Gavazzi2000}{G+00}
\defcitealias{Osorno2023}{O+23}

\title[Dynamics and Ionization in M87's filaments]{Observations of AGN-driven feedback: dynamics and ionization of the filaments in M87}

\author[C. Poitras et al.]{Camille Poitras$^{1}$\thanks{E-mail: camille.poitras.2@ulaval.ca}, 
Marie-Lou Gendron-Marsolais$^{1}$,
Valeria Olivares$^{2,3}$,
Yuan Li$^{4}$,
Adrien Picquenot$^{5}$,\newauthor
Aurora Simionescu$^{6}$,
Matteo Fossati$^{7}$,
Alessandro Boselli$^{8}$,
Laura Hermosa Muñoz$^{9}$,
Sara Cazzoli$^{10}$,\newauthor
Julie Hlavacek-Larrondo$^{11}$,
Annabelle Richard-Laferrière$^{12}$,
\\
$^{1}$D\'epartement de physique, de g\'enie physique et d'optique, Universit\'e Laval, Qu\'ebec (QC), G1V 0A6, Canada\\
$^{2}$Departamento de Física, Universidad de Santiago de Chile, Santiago, Chile\\
$^{3}$Center for Interdisciplinary Research in Astrophysics and Space Exploration (CIRAS), Universidad de Santiago de Chile, Santiago 9170124, Chile\\
$^{4}$Department of Astronomy, University of Massachusetts, Amherst, MA 01003, USA\\
$^{5}$Department of Astronomy, Louisiana State University, LA, USA\\
$^{6}$SRON Space Research Organization Netherlands, Niels Bohrweg 4, 2333CA Leiden, The Netherlands\\
$^{7}$Dipartimento di Fisica G. Occhialini, Università degli Studi di Milano-Bicocca, Piazza della Scienza 3, 20126 Milano, Italy\\
$^{8}$Aix-Marseille Université, CNRS, CNES, LAM, Marseille F-13388, France\\
$^{9}$Centro de Astrobiología (CAB), CSIC-INTA, Camino Bajo del Castillo s/n, E-28692 Villanueva de la Cañada, Madrid, Spain\\
$^{10}$Instituto de Astrofísica de Andalucía, IAA-CSIC, Glorieta de la Astronomía S/N, Granada 18008, Spain\\
$^{11}$D\'epartement de Physique, Universit\'e de Montr\'eal, Succ. Centre-Ville, Montr\'eal, Qu\'ebec, H3C 3J7, Canada\\
$^{12}$Institute of Astronomy, University of Cambridge, Madingley Road, Cambridge CB3 0HA, UK
}

\date{Accepted 2025 November 06. Received 2025 November 06; in original form 2025 September 04.}

\pubyear{\the\year{}}

\begin{document}
\label{firstpage}
\pagerange{\pageref{firstpage}--\pageref{lastpage}}
\maketitle

\begin{abstract}
    We present a comprehensive kinematic and ionization analysis of the warm ionized filaments ($\sim$10$^4$~K) in M87, the central galaxy of the Virgo cluster, using new integral field spectroscopy from MEGARA (GTC) and SITELLE (CFHT). MEGARA targets the southeastern (SE) filaments ($\sim$3~kpc from the nucleus), coincident with the only known molecular gas clump, and the far eastern (FE) filament ($\sim$15~kpc), spatially isolated within an old radio lobe. SITELLE fully maps the filaments, offering the first complete views of their kinematics and excitation. Combined with archival ALMA, MUSE and \textit{Chandra} data, these observations offer a multi-phase view of gas dynamics. The filaments display complex motions inconsistent with simple rotation. Velocity structure functions (VSFs) of the warm and cold gas in the central and SE filaments show consistent steep slopes (> 2/3) and flattening on small scales of a few hundred parsecs, possibly suggesting energy injection from Type Ia supernovae, though interpretation is method-limited. The FE filament shows lower VSF amplitude, suggesting less active driving. ALMA CO emission is co-spatial and kinematically aligned with the ionized gas, the latter showing broader velocity dispersions. Ionization diagnostics indicate AGN-related processes (e.g., shocks) dominate, with higher-energy excitation near the radio lobes and (lower-energy) fossil feedback signatures in the FE filament. Finally, the filaments follow the same strong H$\alpha$ - X-ray surface brightness correlation seen in other clusters, supporting local thermal coupling between phases. However, the FE filament deviates from this trend, possibly due to uplift from past AGN outbursts or limitations in the analysis method.
\end{abstract}




\section{Introduction}
\label{sec:introduction}
    In cool-core galaxy clusters, where the central cooling time of the intracluster medium (ICM) is shorter than the Hubble time, the hot X-ray emitting plasma is expected to cool and condense. However, observed cold gas masses and star formation rates fall well below classical cooling flow predictions (see e.g., \citealt{Peterson2006} for a review), although some studies suggest that significant cooling may remain hidden from direct detection (e.g., \citealt{Fabian2022}). This discrepancy is now attributed to AGN-driven feedback, where mechanical and radiative processes heat the ICM, limiting condensation into a multiphase medium (e.g., \citealt{McNamara2007} and \citealt{Hlavacek-Larrondo2022} for reviews; \citealt{Gaspari2017b}). Nevertheless, how AGN energy is transported, dissipated, and coupled to different gas phases remains unclear.

    Warm ionized filaments ($\sim10^4$~K) are common in brightest cluster galaxies (BCGs) and are believed to trace the dynamic interplay between AGN feedback and the ICM (e.g., \citealt{Gaspari2018, Li2020, Olivares2023}). Extending tens of kiloparsecs, these structures are typically co-spatial with soft X-ray-emitting plasma ($\sim10^7$–$10^8$~K) and cold molecular gas ($\sim$10–100~K), and often surround radio lobes (e.g., \citealt{Salome2008, Russell2017, Olivares2022, Olivares2022b, Olivares2025}). Despite extensive observations, their origin remains debated. One leading scenario suggests that the filaments condense \textit{in situ} from the ICM via thermal instabilities. In several models, this is triggered by the uplift of low-entropy gas by AGN-inflated bubbles to altitudes where the gas becomes thermally unstable, leading to precipitation that can rain back toward the center, triggering the supermassive black hole (SMBH; e.g., \citealt{Pizzolato2005, McCourt2012, Gaspari2013, Li2015, Voit2015, McNamara2016, Voit2017}). Alternatively, external processes such as gas-rich mergers or ram-pressure stripping may deliver multiphase material to the cluster core \citep{Sparks1993, Weil1997, Mayer2006}.

    M87 (Virgo A, NGC 4486, 3C274), the central galaxy of the nearby Virgo Cluster ($\sim16.5$~Mpc; \citealt{Cantiello2024}), provides a well-resolved environment for studying these processes. It hosts a $\sim6.5\times10^9$M$_\odot$ SMBH \citep{EHT2019}, powerful relativistic jets with extended radio lobes \citep{Hines1989, Owen2000}, and complex X-ray structures, including elongated hot gas filaments within a diffuse halo (e.g., \citealt{Young2002, Forman2007, Churazov2008, Werner2010}). M87 is also surrounded by a prominent multiphase filamentary nebula first identified decades ago \citep{arp1967counter, Ford1979, Sparks1993}. The warm ionized component extends out to $\sim18$~kpc from the nucleus, exhibiting complex morphologies that closely trace buoyantly rising X-ray cavities and radio structures \citep[hereafter \citetalias{Boselli2019}]{Boselli2019}, with co-spatial UV filaments of similar extent \citep{Sparks2009, Tamhane2025}. Notably, M87 is unusually deficient in molecular gas, hosting only a compact CO-bright clump $\sim$3~kpc southeast of the nucleus \citep{Simionescu2018}. Moreover, some filaments appear detached from the nebulae (\citealt{Gavazzi2000}, hereafter \citetalias{Gavazzi2000}; \citetalias{Boselli2019}) and current AGN activity. These may preserve signatures of past feedback events and offer unique insight into the long-term impacts of AGN-driven outbursts.
    
    Despite a rich history of multi-wavelength studies (e.g., \citealt{Sparks1993, Sparks2004, Churazov2001, Churazov2008, Young2002, Forman2007, Forman2017, Werner2013, Simionescu2018, Boselli2019}), a comprehensive dynamical characterization of M87's warm ionized filament network remains incomplete. As a result, fundamental questions persist regarding filament formation, motion, and mechanisms by which they trace or respond to both current and past AGN feedback.

    In this context, we present new integral field spectroscopy of two key regions in M87’s filamentary nebula with MEGARA. The first targets the inner region ($\sim$3~kpc southeast of the nucleus) beyond the inner radio lobes and overleaping the only known molecular gas clump. This region lies inside and near the edge of a shock cocoon propagating through M87’s X-ray atmosphere \citep{Million2010, Simionescu2018}. The second probes the outer eastern filaments ($\sim$15 kpc from the nucleus), spatially isolated within an old lobe. Both regions fall outside the previous MUSE wide field mode (WFM) pointing (1$\arcmin~\times~1\arcmin$). Complementary SITELLE imaging Fourier transform spectroscopy provides full mapping of the ionized gas across the network. Our aims are to: (1) characterize ionized gas kinematics and turbulence via velocity structure functions (VSFs) and assess coupling with molecular gas; (2) perform emission line diagnostics; and (3) revisit the H$\alpha$ – X-ray spatial correlation at large radii, building on \citet{Olivares2025}. 
    
    This paper is structured as follows: Section~\ref{sec:observations_reduction} describes MEGARA and SITELLE observations, data reduction, and archival datasets. Results addressing our three main goals are presented and discussed in sections \ref{sec:kinematics} through \ref{sec:Ha-RX_correlation}, with conclusions in Section~\ref{sec:conclusions}. For consistency with previous studies \citep{Werner2013, Simionescu2018}, we adopt a systemic velocity of $v = 1307$~km~s$^{-1}$ for M87 ($z = 0.00436$,  $D_L = 16.1$~Mpc; \citealt{Blakeslee2009}), where 1\arcmin = 4.68~kpc.

\section{Observations and Data Reduction}
\label{sec:observations_reduction}
\begin{figure*}
            \centering
            \includegraphics[width=\textwidth]{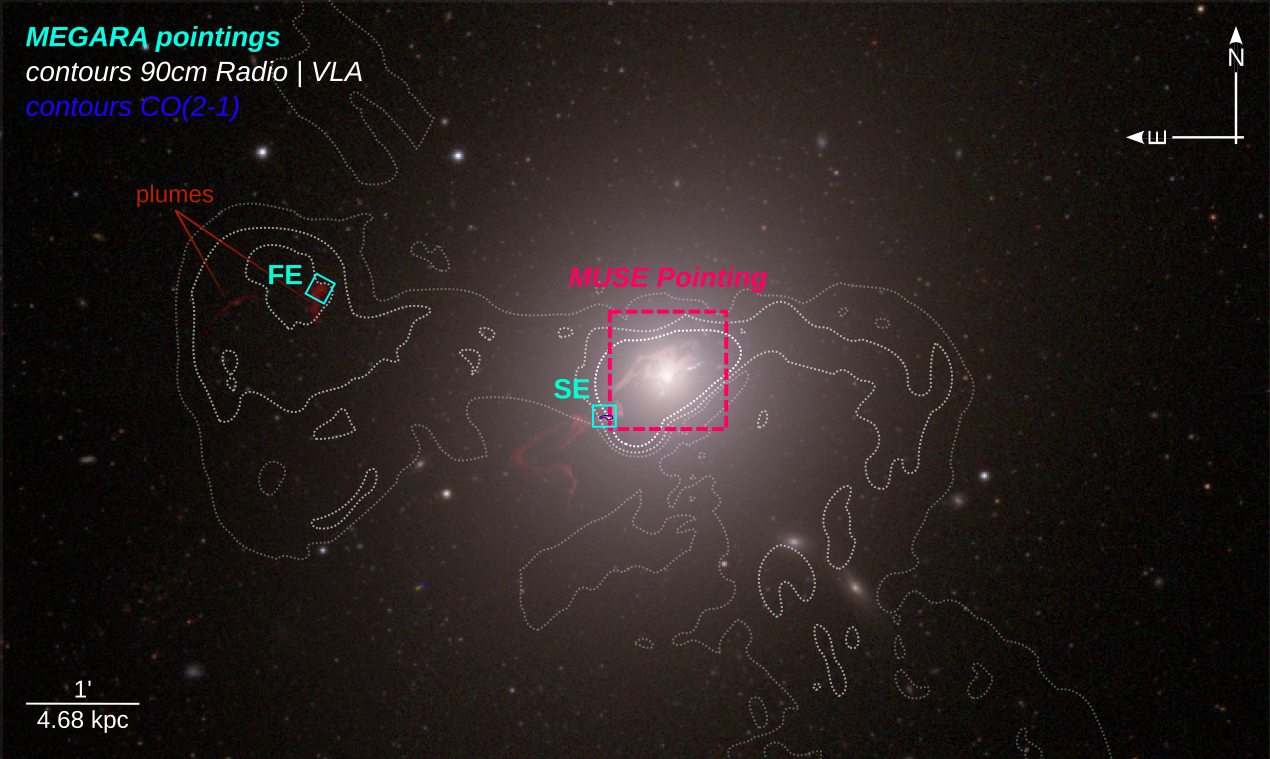}
            \caption{\small Composite image of M87 combining SDSS optical data \citep{SDSS2022} with VESTIGE H$\alpha$ + [\ion{N}{ii}] emission (in red), obtained with MegaCam/CFHT \citep{Boselli2018, Boselli2019}. The image reveals the central filamentary network and two isolated plumes extending northeastward at projected distances of $\sim$15~kpc and $\sim$18~kpc from the galaxy center. Previous MUSE WFM observations is outline in pink, the MEGARA pointings toward the SE and FE filaments are shown in cyan. The blue contour highlights the only known CO(2-1) molecular gas, detected in the SE filaments. Dotted grey contours indicate the 90~cm VLA radio continuum \citep{Owen2000}. The field correspond to the full width and 60\% of the height of the SITELLE FOV, centered on the galaxy.}
             \label{fig:M87Filaments}
        \end{figure*}

    This study is based on a comprehensive dataset, incorporating both new observations and archival data used for the core analysis. A summary of these data is provided in Table~\ref{tab:DetailsObservations}. Details of the new data and their reduction are described in Section~\ref{obs_megara} and \ref{sec:obs_sitelle}, while archival data are documented in the cited references. An overview of M87, its warm ionized filaments, and the main instrument pointings used in this study is shown in Figure~\ref{fig:M87Filaments}.
       
        \begin{table*}
            \centering\small
            \caption{Summary of M87 observations used in our analysis.}
            \label{tab:DetailsObservations}
            \begin{tabular}{lccl}
                \hline
                Facility | Instrument (Mode / Filter)      & Exposure Time    & Prog./Obs. ID      & References$^a$ \\
                \hline
                \textit{CO(2-1)} & & & \\
                    ALMA (Band 6 / 211-275 GHz)            & 5.4~h            & 2013.1.00862.S     & \citet{Simionescu2018} \\
                \hline
                \textit{[\ion{C}{II}]$\micron$} & & & \\
                    Herschel Space Observatory | PACS      & 10,880~s         & 1342236278         & \citet{Werner2013} \\
                \hline
                \textit{Optical} & & & \\
                    GTC | MEGARA (LCB, VPH675-LR)          &                  &                    & \\
                    \quad SE Filaments                     & $\sim2$~h        & GTCMULTIPLE2H-23B  & * \\
                    \vspace{0.4em}
                    \quad FE Filament                      & $\sim5$~h        & ...                & * \\
                    CFHT | SITELLE                         &                  &                    & \\
                    \quad (SN1)                            & $\sim1.84$~h     & 2388234            & * \\
                    \vspace{0.4em}
                    \quad (SN3)                            & $\sim5.67$~h     & 2274329            & * \\
                    VLT | MUSE (WFM)                       & 2 $\times$ 0.5~h & 60.A-9312          & \cite{Boselli2019} \\
                \hline
                \textit{X-Ray (0.5-2.0~keV)} & & & \\
                    Chandra                                & 1745 ks          & see \citet{Olivares2025}, Supp. Info & e.g., \citet{Young2002, Forman2005, Forman2007} \\
                \hline
                \multicolumn{4}{p{0.95\textwidth}}{\footnotesize * This paper.} \\
                \multicolumn{4}{p{0.95\textwidth}}{\footnotesize $^a$ Reference(s) to the paper in which the mentioned data used in this study -- or part of it, in the case of X-ray -- were first published.} \\
            \end{tabular}
        \end{table*}

    \subsection{MEGARA}
    \label{obs_megara}
        The observations were conducted in February 2024 (PI: M. Gendron-Marsolais) using the optical integral field unit (IFU) MEGARA in LCB mode. MEGARA is installed on the \textit{Gran Telescopio Canarias} (GTC), a 10.4 meter telescope at the Roque de los Muchachos Observatory. The IFU consists of 567 fibers (each with a diameter of 100 \micron) arranged in a square microlens array, covering a field of view (FOV) of 12.5$\arcsec$~$\times$~11.3$\arcsec$. Each microlens has a hexagonal shape inscribed within a projected 0.62$\arcsec$ circular diameter. For these observations, LR-R Volume-Phased Holographic (VPH, i.e., VPH675-LR) was used, covering 6096.5–7303.2~\si{\angstrom} with a spectral resolution of $R \sim 6000$, corresponding to a velocity resolution of $\Delta v \sim $50~km~s$^{-1}$. The average seeing was 1.0$\arcsec$ ($\sim$78~pc).

        Two regions in M87 were targeted, sampling different ionized-gas environments. The southeastern filaments (\textit{SE filaments}, $\sim$3~kpc from the nucleus) form a knotty complex and host the only known detection of molecular gas in M87 \citep{Simionescu2018}. The far eastern filament (\textit{FE filament}, $\sim$15~kpc), the ‘plume’ first identified by \citetalias{Gavazzi2000}, is completely isolated from the main nebula. The SE filaments were observed in three observing blocks during the same night, while the FE filament was observed in two. The FE pointing was rotated by $60\degr$ northeast to match the filament's orientation.

        Each observing block was reduced independently using the MEGARA Data Reduction Pipeline \texttt{MEGARA DRP}\footnote{\url{https://github.com/guaix-ucm/megaradrp}} (version 0.15). The pipeline performs bias subtraction, flat-field correction, spectra tracing and extraction, fiber and pixel transmission correction as well as wavelength and flux calibration.

        During observations, MEGARA simultaneously uses 56 dedicated sky-subtraction fibers arranged in eight groups located 1.75-2.0\arcmin from the FOV center. While the \texttt{MEGARA DRP} performs sky subtraction, significant residuals remained after processing. These arise because sky spectra vary with fiber positions in the pseudo-slits, which correspond to their locations at the spectrograph entrance. To mitigate these residuals, we implemented an alternative sky subtraction method. Instead of subtracting a single mean sky spectrum from all fibers, we subtracted the mean of the spectra from the two nearest sky fiber groups for each fiber within the pseudo-slit. This approach significantly improved the sky subtraction quality, leaving only minor residuals generally outside the emission lines of interest. We also found pixel-to-pixel shifts up to 0.71~\si{\angstrom} in the central position of the [\ion{O}{I}]$\lambda$6300 sky line, likely due to inadequate wavelength calibration. Following \cite{arroyo2024unraveling}, we treated the [\ion{O}{I}]$\lambda$6300 sky line as a reference: for each spectrum, we fit it with a single Gaussian to determine its observed position, then subtracted the offset from the entire spectrum. This approach effectively minimizes wavelength calibration uncertainties, ensuring accurate kinematics.

        Finally, each observing block was resampled onto a regularized grid with 0.31$\arcsec$ square spaxels, producing data cubes with dimension of 43~$\times$~40~$\times$~4300, corresponding to a total of 1720 spectra per cube. The final two data cubes were obtained by combining the individual cubes from their respective observing blocks, using a median and ensuring the proper alignment of their central positions.

        \subsubsection{Stellar Continuum Modeling and Subtraction}
        \label{sec:stellar_cont_megara}
            Accurate stellar continuum subtraction is essential to isolate ionized gas emission, especially in early-type galaxies like M87 where stellar absorption dominates the central regions. Although the SE filaments lie at a moderate distance from the center, measurable stellar absorption features (e.g., H$\alpha$ and \ion{Fe}{I}$\lambda$6495) remain and must be corrected. In contrast, the FE filament exhibits little to no stellar absorption, so continuum subtraction was applied only to the SE data cube.

            To measure and remove the stellar continuum, we constructed a representative median spectrum by averaging all spaxels within the FOV, excluding edge regions prone to instrumental artifacts. This was justified by the uniform continuum shape and consistent depth of the stellar absorption lines across the field, in particular the isolated \ion{Fe}{I}$\lambda$6495 line free from nebular emission. This approach assumes negligible stellar age or metallicity gradients in the modelled region. The median spectrum was fitted using pPXF algorithm (Penalized Pixel-Fitting; \citealt{Cappellari2004}). During the fit, we masked emission lines, strong sky residuals, and telluric bands to avoid contamination. We used MILES stellar templates \citep{Vazdekis2010}, based on Padova+00 evolutionary tracks \citep{Sánchez-Blázquez2014, Falcón-Barroso2011}, spanning 3525-7500~\si{\angstrom} at 2.5~\si{\angstrom} full width at half maximum (FWHM) resolution. Although MEGARA provides a higher spectral resolution ($R \sim$ 6000), this mismatch is not critical in our case, as the intrinsic stellar velocity dispersion in M87 ($\sim$250~km~s$^{-1}$ between 2-12~kpc, decreasing by only $\sim$10\% at larger radii; \citealt{Liepold2023}) broadens the absorption lines well beyond the template resolution. The best-fit pPXF model was interpolated onto the spectral grid of each spaxel. The resulting fit to the median spectrum is shown in Appendix~\ref{appendix:stellar_continuum_spectra}. To recover the absolute continuum level, the model was rescaled individually for each spaxel to match its mean observed flux. The final data cube was finally corrected by subtracting the resulting stellar absorption spectrum from each spaxel.

        \subsubsection{Emission Line Fitting}
        \label{sec:emission_line_fitting}
            For each spectrum, the H$\alpha$, [\ion{N}{II}]$\lambda$6548,~6583 and [\ion{S}{II}]$\lambda$6716,~6731 lines were fitted simultaneously with Gaussian profiles using the \texttt{LMFIT} routine, based on the Levenberg-Marquardt least-squares algorithm \citep{Newville2016}. The fits were constrained to share a common velocity shift and line broadening. The [\ion{O}{I}]$\lambda$6300,~6364 lines were excluded from the fit, as they are too faint to be measured reliably in individual spaxels. Nonetheless, in integrated spectra combining spaxels with H$\alpha$ signal-to-noise ratio (SNR) greater than 3, [\ion{O}{I}]$\lambda$6300 emission is more clearly detected, with surface brightness of $\sim$2.9~$\times$~10$^{-17}$~erg~s$^{-1}$~cm$^{-2}$~arcsec$^{-2}$ and $\sim$5.5~$\times$~10$^{-17}$~erg~s$^{-1}$~cm$^{-2}$~arcsec$^{-2}$ for the SE and FE filaments, respectively. In both integrated spectra, the line's velocity and velocity dispersion are consistent with those of the other emission lines. 

\begin{figure*}
                    \centering
                    \includegraphics[width=\textwidth]{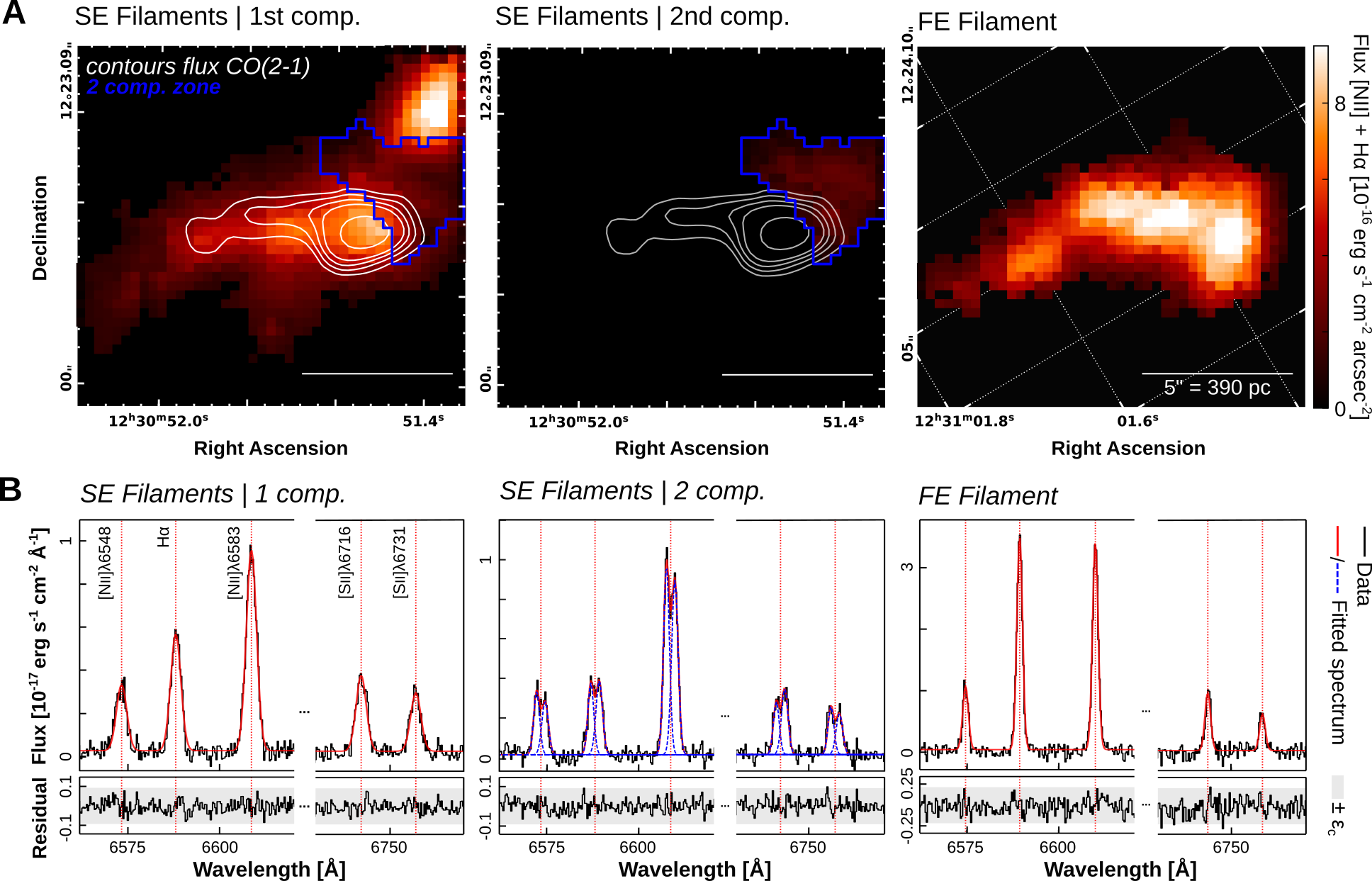}
                    \caption{\small (A) MEGARA emission maps of [\ion{N}{II}]$\lambda$6583 + H$\alpha$ for the first and second kinematic components of the SE filaments (continuum-subtracted), with North at the top and East to the left, as well as for the FE filament. The blue-shaded region outlines the area where the presence of two components if confirmed. Only spaxels with SNR$_{\text{H}\alpha} > 3$ are shown. CO(2-1) flux contours from ALMA (above the 3$\sigma$ level) are overlaid on the SE filament maps, ranging from $\sim$1 to $\sim$6~mJy~beam$^{-1}$. All scale bars correspond to 5$\arcsec$. (B) Representative MEGARA spectra from selected spaxels: from left to right, one from the SE filament with a single component, one with two components, and one from the FE filament. A broken axis has been applied to better visualize the main spectral lines, which are marked by red dotted lines. For each spectrum, the fitted line profile (in red) is shown, along with individual components where applicable (blue dashed curves). The bottom panel displays the residuals of the fits, with the shaded region indicating the $\pm \epsilon_c$ interval.}
                     \label{fig:NIIHaMaps-Spectra}
                \end{figure*}

            To fit the emission lines, we allowed for the possibility of multiple kinematic components. A preliminary visual inspection revealed that parts of the SE filaments contain two kinematically distinct components with similar velocity dispersions. To optimize the fitting process and determine the appropriate number of components, a progressive method was employed. Initially, all spectra were fitted with a single component. To assess whether an additional component was required, we applied the statistical method proposed by \citet{Cazzoli2018}. This method involves measuring the standard deviation of a line-free continuum region $\epsilon_c$ (in this case, between [\ion{N}{II}]$\lambda$6583 and [\ion{S}{II}]$\lambda$6716), and comparing it to the standard deviation of the residuals between the fitted and observed emission lines $\epsilon_\text{line}$. If $\epsilon_\text{line} > 3 \times \epsilon_c$ (following a $3\sigma$ criterion), the addition of a second component is considered. A third component was also tested but never required. A visual inspection of the spectra was also performed to validate the statistical selection of components. Spectra exhibiting two distinct components are primarily located in the northwestern region of the SE filaments' FOV (see Figure~\ref{fig:NIIHaMaps-Spectra}). Additionally, a slight asymmetry in the optical emission line profiles -- located near the eastern edge of the CO(2-1) detection -- hints at a second component there, though this is uncertain: Gaussian fitting becomes ambiguous when components strongly overlap, producing multiple possible solutions. As a result, two-component fits were only retained when statistically justified, exclusively in the northwestern region. In the surrounding areas, transitions between one and two components appear slightly abrupt, an effect likely reflecting the intrinsic limitations of Gaussian fitting.

            Finally, the number of components, denoted by $N$, was statistically validated using the Akaike Information Criterion (AIC) \citep{akaike2011akaike}. In most cases, the two-component fits in the northwestern region are justified when the AIC difference, defined as $\Delta \text{AIC} = \text{AIC}_{N + 1} - \text{AIC}_N$, exceeds 10 \citep{Wei2016}.

            For the remainder of the SE filaments analysis, we define the first component as the one with velocities between approximately -170 to -75~km~s$^{-1}$, which is dynamically connected to the rest of the filament where only a single component is detected. The second component corresponds to velocities within approximately -75 to -30~km~s$^{-1}$. Throughout this paper, we use the label \textit{1 comp.} to refer to the maps from single-component fits, including those within the two-component region. \textit{1st comp.} denotes the first component within the two-component region and single-component spaxels elsewhere. \textit{2nd comp.} refers exclusively to the second component in the two-component region.

        \subsubsection{Map Generation}
        \label{sec:megara_map_generation}
            For each fitted emission line, we measured the central wavelength, velocity dispersion, and flux, along with their uncertainties. The uncertainties from the fitting procedure, which generally exceed those from instrumental calibration, were fully accounted for. These parameters were used to produce flux, velocity, and velocity dispersion maps. Fluxes were corrected for Galactic extinction assuming $A_V = 0.063$~mag (NASA/IPAC Extragalactic Database\footnote{\url{https://ned.ipac.caltech.edu/}}) and $R_V = 3.1$, following the \citet{Cardelli1989} extinction law, with a wavelength-dependent correction applied to each line. Figure~\ref{fig:NIIHaMaps-Spectra} shows the [\ion{N}{II}] + H$\alpha$\footnote{[\ion{N}{II}]$\lambda$6583 + H$\alpha$; this definition will be used throughout the paper.} maps along with examples of fitted spectra. Radial velocities were corrected for barycentric motion. Since the observed velocity dispersions $\sigma_\text{obs}$ were consistently broader than the instrumental width $\sigma_\text{ins}$, estimated from the width of the [\ion{O}{I}]$\lambda$6300 sky line, a correction for instrumental broadening was applied using the following relation
            
                \begin{equation}
                    \sigma_\text{line} = \sqrt{\sigma_\text{obs}^2 - \sigma_\text{ins}^2} \;.
                \end{equation}

    \subsection{SITELLE}
    \label{sec:obs_sitelle}
        M87 was observed with the SITELLE imaging Fourier transform spectrometer (iFTS), installed on the 3.6-meter Canada-France-Hawaii Telescope (CFHT), as a part of two observing programs (PI: Julie Hlavacek-Larrondo, 2018; PI: Annabelle Richard-Laferrière, 2019). SITELLE provides a 11\arcmin~$\times$~11\arcmin FOV, covering the entire filamentary network and its surroundings, with spatial sampling of 0.32$\arcsec$ per spaxel \citep{Drissen2019}. The 2018 observations used the SN3 filter ($R \sim 950$, 651-685~nm), targeting H$\alpha$, [\ion{N}{II}]$\lambda$6548,~6583 and [\ion{S}{II}]$\lambda$6716,~6731 emission lines. The typical seeing during these observations was $\sim1.2\arcsec$. In 2019, additional observations were obtained using the SN1 filter ($R \sim 680$, 365-385~nm), which covers the [\ion{O}{II}]$\lambda\lambda$3727-3729 doublet (unresolved). 

        Data reduction used SITELLE's dedicated pipelines, \texttt{ORBS} and \texttt{ORCS}\footnote{\url{https://github.com/thomasorb/orbs}} (version 3.0; \citealt{Martin2021}). Wavelength calibration was initially carried out with a high-resolution laser source, but instrument flexures during target acquisition can introduce velocity shifts of $\sim$15-20~km~s$^{-1}$ across the FOV, with variations in M87's central region remaining below 5~km~s$^{-1}$. These were corrected by measuring OH sky line central positions within cubes, following \citet{Martin2018}. Flux calibration was performed for each filter using stars in the field matched to \textit{Gaia Data Release 3} \citep{Gaia2023} sources with calibrated optical spectra. Synthetic photometry was derived by integrating Gaia spectra over the filter transmission curve, then compared to SITELLE fluxes measured from 2D Gaussian fits of stellar profiles in deep frames (i.e., the sum of all interferogram frames). For the SN3 filter, a linear regression between SITELLE and Gaia-based fluxes yielded a slope of $0.942 \pm 0.092$, indicating that SITELLE fluxes slightly underestimate the true flux. A correction factor of 1/0.942 was therefore applied. For the SN1 filter, a larger discrepancy was found, with a slope of $0.804 \pm 0.040$ and the corresponding factor was applied accordingly. A detailed description of this type of calibration procedure for SITELLE will be presented in Vicens-Mouret (in prep).

        Prior to the emission line fitting, sky subtraction was performed using a circular region with a radius of 0.5\arcmin, centered at ($\alpha$:~12h31m08.0636s, $\delta$:~+12h24m59.761s). This area contains neither point sources nor detectable source emission and provides a representative sky background comparable to other object-free regions of the FOV.
        
        Stellar continuum subtraction for SITELLE required a spatially adaptive approach due to the noticeable spectral variations across the field. Visual inspection of the SN1 and SN3 data cubes revealed significant changes in the continuum level and absorption line depth with galactocentric radius, while stellar absorption velocities remained uniform in both filters, consistent with M87's low rotation ($\sim 25$~km~s$^{-1}$ beyond 5~kpc, decreasing inward; \citealt{Liepold2023}). To account for the spatial variations, the field was divided into concentric annuli centered on the galactic nucleus. Radial boundaries were selected to reflect observed spectral gradients. Within each annulus, spaxels representative of the stellar population were identified based on their low H$\alpha$ emission (< $5~\times~10^{-19}$~erg~s$^{-1}$~cm$^{-2}$) using a flux map derived from the original data cube. Foreground stars and spaxels including the jet were manually excluded. Each annulus contained at least 600 qualifying spaxels. To construct a representative stellar spectrum for each annulus, the selected spectra were first normalized by their mean flux to prevent bias from bright spaxels, then median-combined. The resulting spectra were fitted with pPXF using MILES templates, fitting the SN1 and SN3 filter simultaneously. The best-fit models were linearly interpolated to the observed spectral grid of each filter and rescaled to match each pixel’s mean flux for the continuum level. Subtracting the appropriately scaled model for each annulus produced the final continuum-corrected cube. An example integrated spectrum and its fitted model are shown in Appendix~\ref{appendix:stellar_continuum_spectra}.
        
\begin{figure*}
                \centering
                \includegraphics[width=\textwidth]{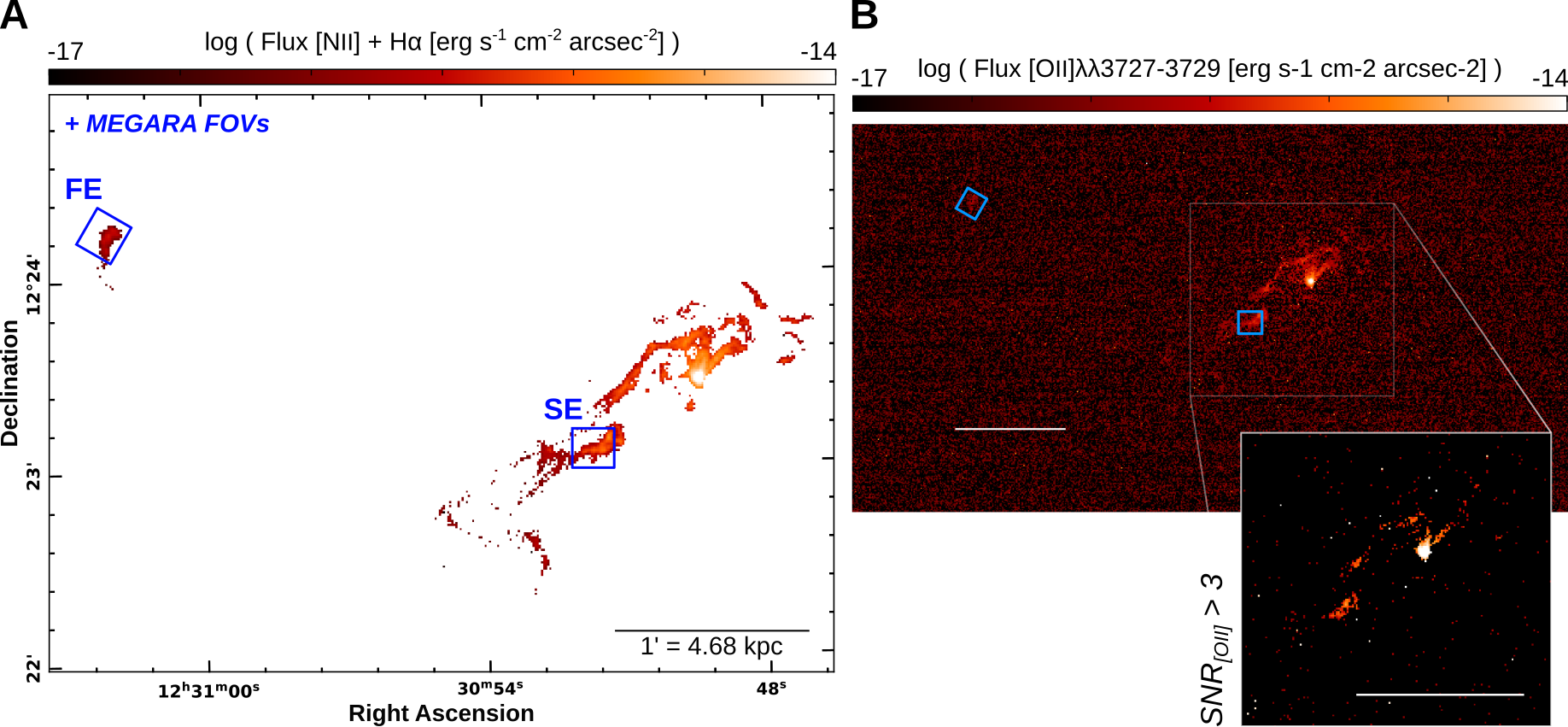}
                \caption{\small (A) SITELLE continuum-subtracted [\ion{N}{II}] + H$\alpha$ emission map showing only spaxels with SNR$_{\text{H}\alpha} > 3$ and velocity error $\leq$~30~km~s$^{-1}$. (B)~Corresponding [\ion{O}{II}]$\lambda\lambda$3727-3729 emission map. The main panel displays all spaxels, while the inset shows a zoomed-in view of the central region with only spaxels satisfying SNR$_{\text{[OII]}} > 3$. In both panels, the MEGARA FOVs targeting the SE and FE filaments are outlined in blue. The displayed maps represent approximately 10\% of SITELLE's full FOV. All scale bars correspond to 1$\arcmin$. Maps are shown with 2~$\times$~2 binning and a logarithmic color scale to enhance contrast between bright and faint emission.}
                 \label{fig:NIIHa-OIIMaps-SITELLE}
            \end{figure*}

        From the calibrated data cubes, with sky and stellar continuum subtracted, \texttt{ORCS} was used to generate maps of the continuum, emission line amplitude and flux, radial velocity, and velocity dispersion, along with their associated uncertainties. To enhance the SNR, maps were spatially binned (2~$\times$~2). A barycentric correction was applied to the velocity map. Fluxes were corrected for Galactic extinction following the method described in Section~\ref{sec:megara_map_generation}. The [\ion{N}{II}] + H$\alpha$ emission map is shown in Figure~\ref{fig:NIIHa-OIIMaps-SITELLE}. Due to the applied SNR$_{\text{H}\alpha} > 3$ threshold, the faint outermost plume filament first identified by \citetalias{Boselli2019} is not visible in the map. The same figure also presents the [\ion{O}{II}]$\lambda\lambda$3727-3729 emission map, where faint [\ion{O}{II}] emission is detected in several filaments, including the FE filament, consistent with earlier reports by \citet{arp1967counter} and \citet{Ford1979}. However, spaxel-by-spaxel SNR remains below 3 across most regions. Combined with the unresolved doublet in SN1 data, this prevents the use of [\ion{O}{II}] for reliable electron density diagnostics. H$\beta$ and [\ion{O}{III}]$\lambda$5007 are outside the SITELLE filter coverage, further limiting ionization diagnostics. Based on the mean SNR of spaxels with SNR$_{\text{H}\alpha}>3$ and the relation from \citet{RousseauNepton2019}, the minimum measurable velocity dispersion is $\sim$180~km~s$^{-1}$ for SN3 and $\sim$375~km~s$^{-1}$ for SN1. Given these limits, velocity-dispersion results are not used in our analysis, though the map is provided for reference in Appendix~\ref{appendix:sitelle_dispersion}.

    \subsection{Archival Data, WCS Alignment and Spatial Resampling}
    \label{sec:wcs_alignment}
        To compare the kinematic and morphological properties of different gas phases, we incorporate archival datasets with our new observations. We directly compare MEGARA maps of the SE filaments with ALMA CO(2-1) flux, velocity and velocity dispersion maps derived via masked-moment analysis (see \citealt{Simionescu2018} for details). We also compare indirectly SITELLE velocity field and with the MUSE velocity map from \citetalias{Boselli2019}, based on the original data of \citet{Emsellem2014}, corrected to the systemic velocity. The MUSE data have a spaxel size of 0.2$\arcsec$, $R\sim1500-3500$, and a seeing of 0.9$\arcsec$. In addition, we directly compare the SITELLE H$\alpha$ map with deep \textit{Chandra} X-ray data (see Table~\ref{tab:DetailsObservations}).

        Spatial consistency was ensured by aligning the World Coordinate Systems (WCS) for all datasets. A visual inspection revealed slight offsets in the MEGARA and MUSE data relative to the others. These were corrected using instrument-specific alignment methods tailored to each dataset and the objectives of the analysis. For MEGARA, which lacks bright point sources in its FOV, alignment was achieved by cross-correlating H$\alpha$ filamentary structures with SITELLE (SNR > 3, unbinned), using the \texttt{correlate2d} function after degrading SITELLE to MEGARA's spaxel grid. The resulting pixel offsets were converted into WCS corrections and applied to MEGARA cubes: ($\alpha$:~–0.000677, $\delta$:~+0.000135)$\degr$ for the SE pointing and ($\alpha$:~–0.000364, $\delta$:~+0.0000854)$\degr$ for the FE pointing. For MUSE, alignment was achieved by matching the galactic center position in its velocity maps to SITELLE, with a final correction of ($\alpha$:~–0.0001785, $\delta$:~+0.0003881)$\degr$. 
        
        To enable direct spaxel-by-spaxel comparisons with ALMA, MEGARA maps were convolved with a Gaussian kernel to match the ALMA beam. The ALMA data reach an rms sensitivity of 0.2~mJy~beam$^{-1}$ per 50~km~s$^{-1}$ channel, with a synthesized beam of 1.4$\arcsec$~$\times$~0.7$\arcsec$ (position angle of 75$\degr$; \citealt{Simionescu2018}). The ALMA data were resampled onto MEGARA’s spaxel grid using \texttt{astropy.reproject} function. This enabled the construction of three key comparative maps: (1) the peak-normalized ([\ion{N}{II}] + H$\alpha$)/CO flux ratio, (2) the line-of-sight (LOS) velocity difference, and (3) the velocity dispersion ratio (see Section~\ref{sec:SE-CO}).

\section{Filament Kinematics}
\label{sec:kinematics}

    \subsection{Velocity Field Across the Filament Network}
    \label{sec:velocity_field_network}
        M87's ionized gas filaments form a complex and extended network whose kinematics have been extensively study for decades. Early long-slit spectroscopy revealed irregular velocities and significant variations across the filaments \citep{Heckman1989, Sparks1993, Werner2013}. In particular, \citet{Sparks1993} proposed a three-dimensional geometry: the northwestern filaments, near the jet, lie in the foreground and are likely outflowing, while the southeastern filaments, near the counter-jet, are in the background and infalling -- a scenario supported by dust morphology and radio lobe orientation.  

\begin{figure}
                \centering
                \includegraphics[width=0.99\columnwidth]{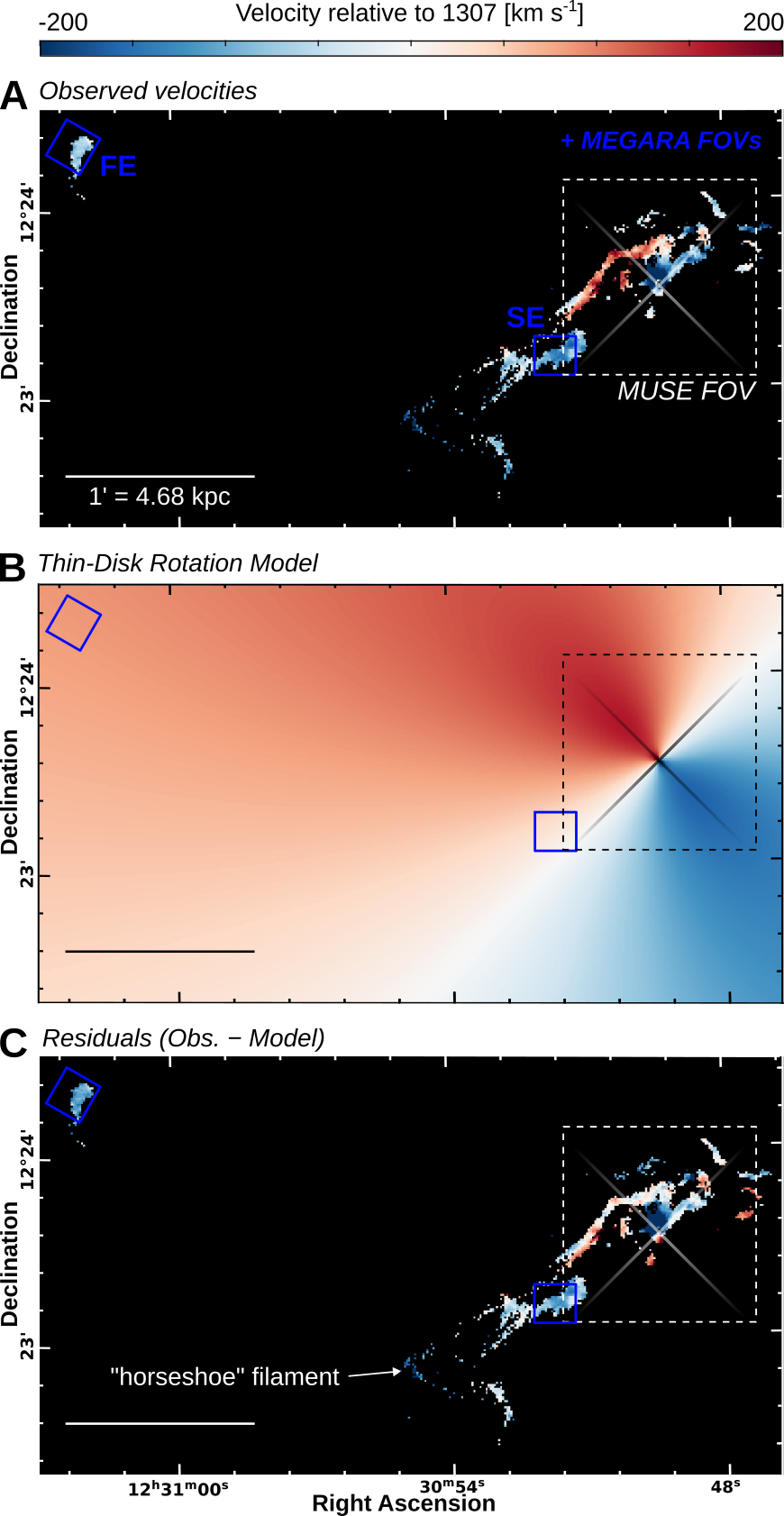}
                \caption{\small (A) SITELLE velocity map showing spaxels with SNR$_{\text{H}\alpha} > 3$ and velocity uncertainties $\leq$ 30~km~s$^{-1}$ (2~$\times$~2 binning). (B) Rotating thin-disk model computed for the combined gravitational potential of the central SMBH and the stellar bulge, based on the best-fit model from \citetalias{Osorno2023}. (C) Residual map showing the difference between the observed velocities and the rotation model. For each panel, faint solid white/black lines centered on the nucleus mark the disk's kinematic P.A. (45\degr) and its perpendicular axis. MEGARA FOVs targeting the SE and FE filaments are outlined in blue, and the existing MUSE WFM observations is outlined with a white/black dashed box. The color scale is consistent across all panels and each scale bar corresponds to 1$\arcmin$.}
                 \label{fig:Vel-Residual-Maps_SITELLE}
            \end{figure}

        With the advent of integral field spectroscopy, recent studies have refined this picture. MUSE WFM observations \citepalias{Boselli2019} revealed highly disordered velocity fields across the central $\sim1\arcmin~\times~1\arcmin$, consistent with turbulence likely driven by AGN feedback from the jet \citep{Li2020}. At smaller scales, \citet[hereafter \citetalias{Osorno2023}]{Osorno2023} used MUSE narrow field mode (NFM) data to probe the nuclear region in detail, unveiling complex kinematics. Nevertheless, previous velocity studies lacked the spatial coverage to trace motion across the full filament system: MUSE WFM covered only the central filaments, while earlier long-slit data extended slightly farther. With SITELLE and MEGARA, we now access both the full spatial extent of the filamentary network and sufficient spatial resolution, including the southeastern and isolated filaments beyond 40$\arcsec$ ($\sim$3.1~kpc) from the nucleus. Velocity map derived from SITELLE and velocity/dispersion maps from MEGARA are shown in Figure~\ref{fig:Vel-Residual-Maps_SITELLE} and Figure~\ref{fig:Vel-SIgMaps_VSFs_MEG_SIT}, respectively.

        In the nuclear region, a small number of SITELLE spaxels confirm the complex velocity structure reported by \citetalias{Osorno2023}. Their analysis showed that the so-called ionized gas “disk” is not a simple rotating structure, but a mix of rotating material, filamentary emission with distinct bulk velocities, and features crossing the nucleus in projection -- producing irregular velocity gradients exceeding 300~km~s$^{-1}$ on subarcsecond scales. Just beyond the nucleus, extending out to the edge of the MUSE WFM pointing, we find excellent agreement with the LOS velocities presented by \citetalias{Boselli2019} and earlier long-slit studies, with disordered motions typically reaching up to $\pm$300~km~s$^{-1}$.

        In the south, the MEGARA data reveal the detailed velocity field of the SE filaments, including a region with resolved two-component profiles located at the edge of the radio counter-jet. The primary component (1st comp.) has a mean LOS velocity of $\sim -120$~km~s$^{-1}$ and a velocity dispersion of $\sim 55$~km~s$^{-1}$. It is kinematically aligned with cold molecular gas traced by CO(2-1) and [\ion{C}{II}]$\lambda$157$\micron$ emission, both previously reported as co-spatial with the ionized gas \citep{Simionescu2018, Werner2013}. A more detailed comparison of these phases is presented in Section~\ref{sec:SE-CO}. In the FE filament, we measure a mean velocity of $\sim -60$~km~s$^{-1}$, consistent with \citetalias{Gavazzi2000}, and a low velocity dispersion of $\sim 30$~km~s$^{-1}$. The kinematic properties of the SE and FE filaments are further discussed in Section~\ref{sec:SE-FE_VSFs} in the context of velocity structure and turbulence.
        
        Finally, SITELLE allows us to probe the most extended southeastern filaments, which had not been kinematically studied before. These structures appear to skirt buoyant radio bubbles of varying ages rising through M87's hot atmosphere \citep{Forman2007}. These outer filaments exhibit mostly negative velocities between $-$250 to 50~km~s$^{-1}$. To assess whether M87's ionized gas filaments show large-scale rotation, we extend the analysis of \citetalias{Osorno2023} beyond the MUSE WFM pointing. Their study focused on the central filaments ($\sim1\arcmin~\times~1\arcmin$), comparing observed velocities to analytical thin-disk rotation models. They concluded that while the innermost filaments deviate strongly from rotation, outer filaments (beyond $\sim$10$\arcsec$) can reasonably be described by rotation in a disk inclined by $i \approx 25\degr$, particularly influenced by the stellar potential. Here, we investigate whether this pattern persists into the more distant southeastern filaments beyond 40$\arcsec$ with SITELLE, which were not included in their analysis. Using the same analytical framework as \citetalias{Osorno2023}, the rotational velocity $v_\text{rot}$ at radius $r$ is given by 
            \begin{equation}
                v_\text{rot}(r) = \sqrt{\frac{G[M_\bullet + M_\text{bulge}(r)]}{r}} \;,
            \end{equation}

        \noindent where $G$ is the gravitational constant, $M_\bullet = 6.0\times10^9$ M$_\odot$ \citepalias{Osorno2023} is the mass of the SMBH, and 
            \begin{equation}
                M_\text{bulge}(r) = \int_0^r 4\pi r'^2 \Upsilon(r')\nu(r') \text{d}r' \;,
            \end{equation}

        \noindent is the enclosed stellar mass within a radius $r$, with $\nu(r')$ is the V-band luminosity density profile from \citet{Gebhardt2009}, and $\Upsilon(r')$ the stellar $M/L$ ratio fitted by \citet{Liepold2023}. As $\nu(r')$ is non-analytic, the integral is evaluated numerically. To project the rotational velocity into the LOS, we calculate
            \begin{equation}
                v_\text{LOS} = v_\text{rot}(r) \sin i \cos \theta \;,
            \end{equation}

        \noindent where $i = 25\degr$ is the disk inclination, and $\theta$ is the azimuthal angle in the disk plane between the major axis (assumed at PA = $45\degr$), adopting the geometric parameters derived by \citetalias{Osorno2023}. We compare this model to the SITELLE velocities via a residual map ($v_\text{obs} - v_\text{model}$), to highlight spatial regions where the observed ionized gas velocities are consistent with a disk-like rotation and where significant deviations occurs. The resulting residual map is shown in Figure~\ref{fig:Vel-Residual-Maps_SITELLE}. Within the MUSE FOV, residuals match \citetalias{Osorno2023}. Beyond this area, few portions of the filaments follow the general trend expected from disk rotation, with residuals on the order of $\sim\pm$50~km~s$^{-1}$. However, the velocity field is more complex. Notably, the SE and FE filaments, as well as the horseshoe-like filament located at $\sim90$$\arcsec$ southeast from the nucleus, exhibit substantial deviations (> 100~km~s$^{-1}$) from model predictions, suggesting additional local processes (e.g., shocks or dynamical instabilities). In particular, the SE filaments lie just beyond the inner radio lobes, while the FE filament is embedded in an older radio lobe. Additionally, the influence of the biconical outflow identified by \citetalias{Osorno2023} may also contribute to the disturbed motions in the inner filaments. Therefore, a simple rotating disk model is insufficient to explain the full range of filament motions observed in M87.

    \subsection{Velocity Structure Functions}
    \label{sec:vsf}
        VSFs have emerged as a valuable tool for characterizing turbulence in filamentary systems and are increasingly applied across entire FOV to investigate dominant turbulent driving mechanisms. In M87, \citet{Li2020} pioneered the use of VSFs in such optical filaments, using MUSE WFM data of the central region. They identified a driving scale of 1-2~kpc, consistent with the AGN jet and X-ray cavity, linking turbulence to jet-driven activity. Here, we extend this approach by analyzing the nature of filament motions using first-order VSFs ($n~=~1$). As a two-point correlation function, the VSF is defined as

            \begin{equation}
                \langle |\mathbf{v}(\mathbf{l})|^n \rangle =  \langle |[\mathbf{v}(\mathbf{x} + \mathbf{l}) - \mathbf{v}(\mathbf{x})]|^n \rangle \;,
            \end{equation}

        \noindent which quantifies the average absolute velocity difference between two points, $\mathbf{x}$ and $\mathbf{x} + \mathbf{l}$, separated by a LOS vector $\mathbf{l}$. VSFs are typically expressed as a function of the scalar separation $l = |\mathbf{l}|$, providing a statistical description of velocity variations across spatial scales. Mean velocity differences $\langle \Delta v \rangle$ are computed by binning separation distances $l$ in logarithmic space: at small scales, they are manually adjusted to account for data discretization, while the larger scales -- where the sampling is more continuous -- fixed width bins are adopted. 

        To ensure the reliability of the VSFs, spaxels with velocity uncertainties greater than 20~km~s$^{-1}$ were excluded, a threshold above the mean uncertainty that does not significantly affect the results. These higher-uncertainty spaxels lie mostly at filament edges where the SNR is lower. Additionally, due to the finite size of the FOV and structures themselves, the number of point pairs decreases with increasing separation. As a result, bins containing fewer than 20\% of the maximum number of pairs are marked in grey in the VSF plots to indicate reduced statistical significance. Each VSF plot also includes a vertical marker at half the size of the spatial window used, highlighting scales where interpretations become uncertain, and the seeing scale to illustrate potential atmospheric effects. Further discussion of these limitations and the uncertainties they introduce in our interpretation is provided in Section~\ref{sec:limitations_vsf}. For the uncertainties of the mean velocity differences, they were estimated using a combined approach that accounts for both error propagation from individual velocity measurements and statistical variations via bootstrapping.

        To guide VSF interpretation, we overlay reference power-law slopes on the plots. Specifically, a slope of 1/3 is shown as a benchmark for classical Kolmogorov turbulence as expected for an incompressible fluid \citep{Kolmogorov1941}. We also include a steeper slope of 2/3, often matching our measurements and commonly observed in other systems (e.g., \citealt{Li2020, Ganguly2023}). It is also worth noting that for compressible (supersonic) turbulence, the expected slope is approximately 1/2 \citep{Boldyrev1998}. While many VSFs follow a clear power-law scaling at smaller spatial separations, several also exhibit a noticeable flattening at intermediate scales. This flattening indicates the dominant driving scale of turbulence, where turbulent kinetic energy injected at large scales begins transferring to smaller scales before being dissipated by viscosity. We return to this interpretation in the context of specific cases discussed in the results. 

        In the following sections, we investigate the VSF derived from different instruments, substructures, and gas phases (sections \ref{sec:SE-FE_VSFs} to \ref{sec:multiple_vsfs}). A global interpretation is presented in Section~\ref{sec:discussion_vsf}, followed by an application to the turbulent heating rate in Section~\ref{sec:heating_rate}. The limitations inherent to the VSF analysis are discussed in Section~\ref{sec:limitations_vsf}.

        \subsubsection{SE and FE Filaments}
        \label{sec:SE-FE_VSFs}
\begin{figure*}
                    \centering
                    \includegraphics[width=0.91\textwidth]{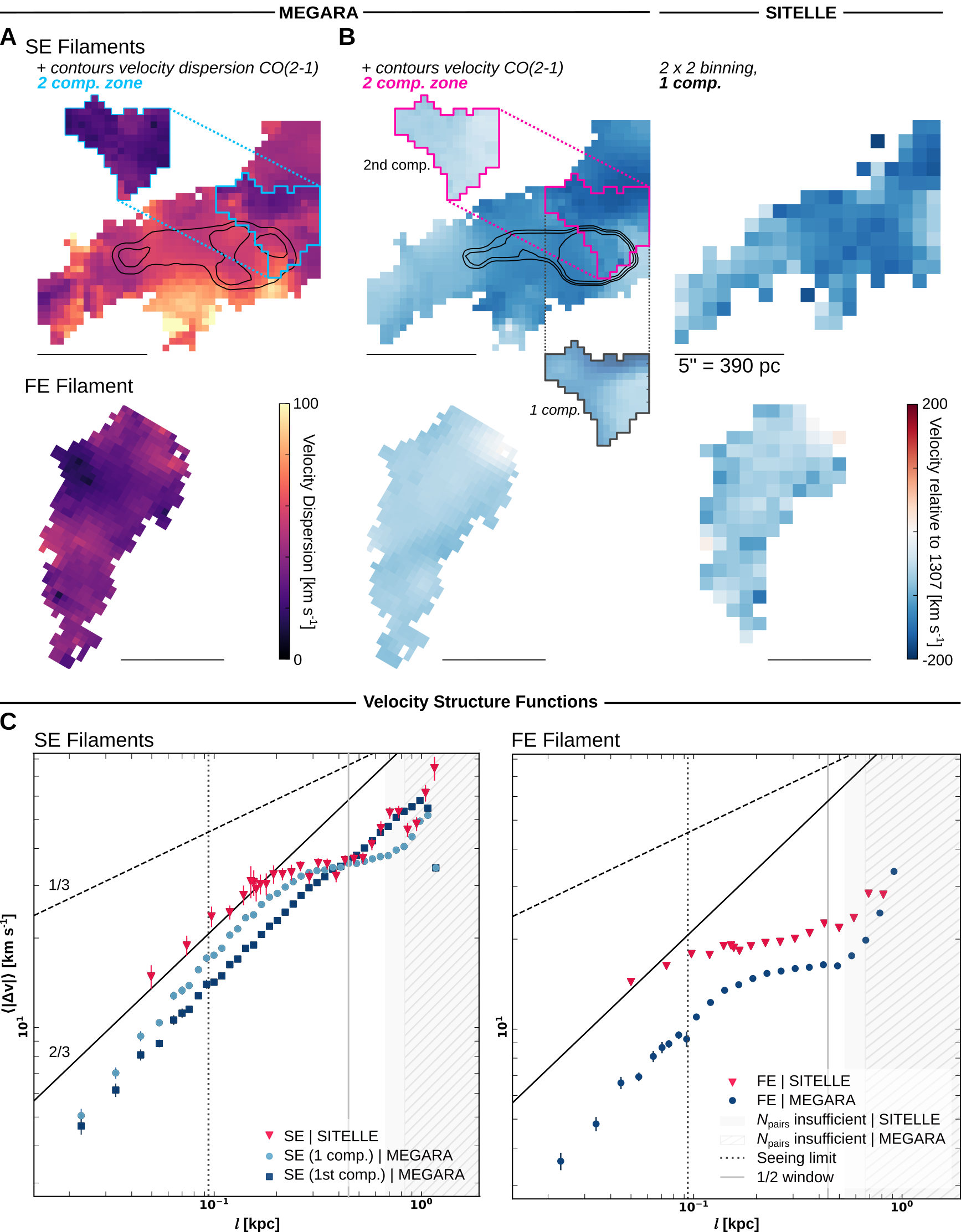}
                    \caption{\small (A) Velocity dispersion maps and (B) velocity maps of the SE (top) and FE (bottom) filaments, obtained with MEGARA (left) and SITELLE (right), showing only spaxels with SNR$_{\text{H}\alpha} > 3$. North is up and East is to the left. For the SE filaments in MEGARA, the main maps display the first component, while separate maps display the second component and, for the velocity map, the single-component fit within the two-component region (outlined). SITELLE maps include only a single velocity component. Color scales are consistent across maps of the same type. All scale bars correspond to 5$\arcsec$. (C) VSFs for both pointings, derived from SITELLE and MEGARA data. Shaded areas indicates bins with an insufficient number of pairs ($< 20\%$ of the maximum pair count). Two reference slopes are shown: 1/3 (Kolmogorov turbulence) and 2/3 (steeper slope). Error bars are displayed, although in some cases they are smaller than the symbol size. The vertical dotted and solid lines indicate the SITELLE seeing limit ($\sim$1.2$\arcsec$) and the half the height of the MEGARA FOV, respectively.}
                     \label{fig:Vel-SIgMaps_VSFs_MEG_SIT}
                \end{figure*}

            VSFs were first computed for the SE and FE filaments using MEGARA velocity maps. An initial set of VSFs was derived under the assumption of a single-component velocity fit for both pointings (1 comp.). For the SE filaments, a second VSF was derived from the primary component (1st comp.) of the two-component fit to assess the effect of overlapping velocity structures. A separate VSF for the second component was not computed due to the limited spatial coverage and its large velocity offset relative to the rest of the filament. MEGARA VSFs were then compared to those from SITELLE, restricted to the same FOVs. The results are shown in Figure~\ref{fig:Vel-SIgMaps_VSFs_MEG_SIT} (panel C).
    
            All VSFs display a clear power-law behavior at small scales, with slopes generally steeper than the Kolmogorov 1/3 value, approaching $\sim$2/3. In the single-component analysis, both MEGARA and SITELLE show a flattening at scales of $\sim$0.2-0.3~kpc. The corresponding velocity amplitudes at the flattening scale are $\sim$30~km~s$^{-1}$ for the SE filaments and $\sim$15–20~km~s$^{-1}$ for the FE filament. At larger scales, the VSF continues to rise, although the declining number of spaxel pairs at large separations limits our ability to draw firm conclusions in this regime.
    
            Interestingly, for the SE filaments, the VSF based on the first component does not exhibit a clear flattening; instead, the power increases steadily toward larger scales. This suggests that the flattening seen in the one component VSF may not correspond to a physical scale break, but rather the impact of LOS complexity. Further discussion of these effects is provided in Section~\ref{sec:limitations_vsf}.
            
            Overall trends are similar between MEGARA and SITELLE, though SITELLE yields slightly higher VSF amplitudes, which may reflect the increased noise of the SITELLE velocity maps in this study (see Section~\ref{sec:limitations_vsf}).
            
        \subsubsection{Comparison Between SE Filaments and CO(2-1)}
        \label{sec:SE-CO}
\begin{figure*}
                    \centering
                    \includegraphics[width=\textwidth]{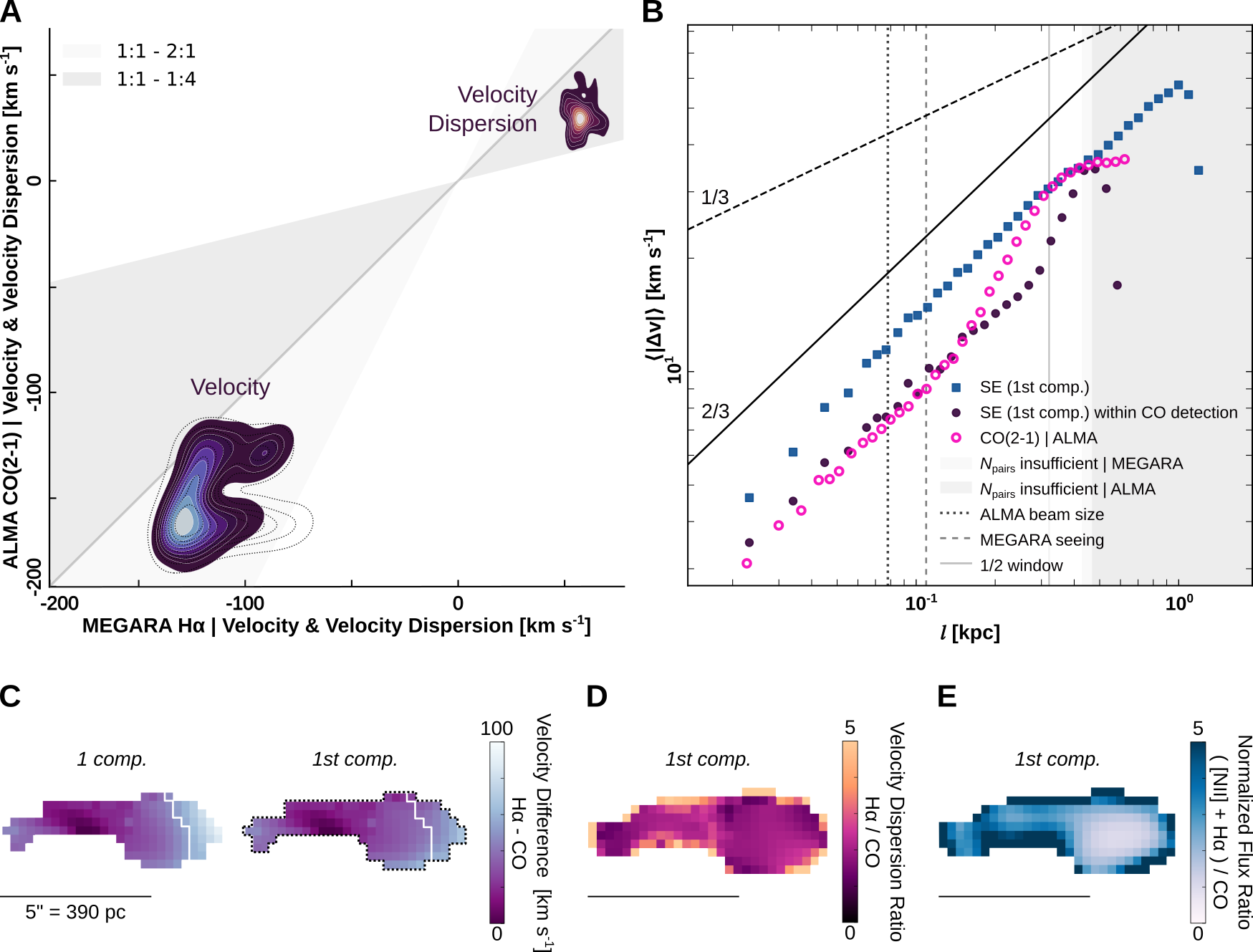}
                    \caption{\small (A) Spaxel-by-spaxel comparison of ALMA CO(2-1) and MEGARA H$\alpha$ (1st comp.) LOS velocities (blue) and velocity dispersion (pink), smoothed using a Gaussian kernel. For the velocity comparison, the dotted contours represent the distribution of spaxels using the 1 comp. fit, illustrating how the use of a two-component decomposition in the appropriate region tightens the overall distribution. Shaded bands indicate regions where the H$\alpha$ values are 1 to 2 times (light grey) and 1 to 4 times (dark grey) the CO values. The 1:1 relation is highlighted by the solid grey line. (B) VSFs for the full SE filament using MEGARA (dark blue), the MEGARA spaxels within the CO(2-1) detection region (black) and the CO(2-1) detection itself (magenta). Two reference slopes are shown: 1/3 (Kolmogorov turbulence) and 2/3 (steeper slope). The vertical dotted, dashed and solid lines indicate the ALMA beam size (1.4$\arcsec$, lager beam axis), the MEGARA seeing limit ($\sim$1$\arcsec$) and the half-width of the CO-detected region (extent from east to west), respectively. (C) Velocity difference map (H$\alpha$ - CO), with the area right of the white line corresponding to the two-component region. (D) Velocity dispersion ratio map (H$\alpha$ / CO); edges should be disregarded. (E)~Normalized ([\ion{N}{II}] + H$\alpha$)/CO flux ratio map. All scale bars correspond to 5$\arcsec$.}
                     \label{fig:ALMA-vs-MEGARA_Maps-VSF}
                \end{figure*}

            To assess whether a spatial and kinematic correspondence exists between the SE ionized filaments observed with MEGARA and the CO(2-1) emission detected by ALMA, we directly compare the two datasets. For this purpose, we use the smoothed and resampled flux, velocity and velocity dispersion maps for both gas phases, along with the corresponding difference and ratio maps as described in Section~\ref{sec:wcs_alignment}. Part of the CO(2-1) emission overlaps the region where MEGARA detected two velocity components. The ALMA observations indicate a LOS velocity of $\sim$130~km~s$^{-1}$, which falls within the range of MEGARA's first component. We therefore use the first component in this region and the single component map elsewhere. All results discussed in this section are presented in Figure~\ref{fig:ALMA-vs-MEGARA_Maps-VSF}. 
            
            Spatial correspondence was first assessed via the normalized flux ratio map of ([\ion{N}{II}] + H$\alpha$)/CO(2–1). The emission peaks of the first component of the SE filament and the CO(2–1) aligned closely, yielding ratios near or slightly below unity. Moving eastward, the ratio increases substantially, with a median value around 4 and even higher values at the edges. Outside the CO detection, strong ionized emission persists without molecular counterpart. This may reflect the destruction or excitation of molecular gas by AGN-driven processes, as suggested by \citet{Simionescu2018}. In contrast, [\ion{C}{II}] emission coincides spatially with the ionized filaments and also extend beyond the CO detection, as reported by \citet{Werner2013}. 
            
            Kinematic correspondence was evaluated through LOS velocity-difference and velocity dispersion ratio maps (H$\alpha$\footnote{H$\alpha$ is used here as a proxy for the ionized gas. All emission lines in a given spectrum share the same kinematic profile in our fits.} / CO) to highlight spatial regions of agreement or discrepancy. We complemented this analysis with spaxel-by-spaxel scatter plots comparing velocity and velocity dispersion between the two tracers, following the approach of \citet{Tremblay2018}.
            
            The velocity-difference map reveals a strong correspondence across most of the overlapping region, with offsets typically below 55 km~s$^{-1}$ and a mean of $(29 \pm 1)$~km~s$^{-1}$. This supports the idea that ionized and molecular gas are largely co-spatial and kinematically aligned. A similar trend is observed in [\ion{C}{II}] emission, which varies from approximately $-100$~km~s$^{-1}$ in the southeast to $-120$~km~s$^{-1}$ northwest along the SE filaments \citep{Werner2013}, roughly consistent with MEGARA H$\alpha$ velocities spanning approximately from $-80$~km~s$^{-1}$ to $-150$~km~s$^{-1}$. For the CO, larger velocity offsets appear at the eastern and western edges of the CO detection. In the east, the discrepancy may reflect the limits of a single-component Gaussian fit, consistent with previous indications of a second velocity component in this region (see Section~\ref{sec:emission_line_fitting}). In the west, using two components reduces the mean offset from approximately 70~km~s$^{-1}$ to 50~km~s$^{-1}$, suggesting a more accurate kinematic decomposition. A spaxel-by-spaxel LOS velocity comparison (see Fig~\ref{fig:ALMA-vs-MEGARA_Maps-VSF}, panel A) confirms the alignment: most points fall near the 1:1 relation, with some dispersion toward 2:1. Notably, the use of two velocity components in MEGARA improves the agreement. Given this consistency, we further compare turbulent properties using VSFs constructed from both velocity maps (i.e., MEGARA, ALMA). On small scales, CO(2–1) and H$\alpha$ VSFs follow similar power-law trends with slopes close to 2/3. When the H$\alpha$ VSF is restricted to the MEGARA spaxels within the CO(2–1) detection footprint, the agreement between the two improves. Within this region, both VSFs show similar power-law slopes and amplitudes at small scales. Beyond $\sim$0.1~kpc, the CO VSF steepens, yielding higher amplitude than H$\alpha$. In contrast, the H$\alpha$ VSF maintains a similar slope to that of the full first-component VSF, remaining roughly parallel. Around 0.2–0.3~kpc -- near the half the width of the CO-emitting region ($\sim$300~pc east–west) -- the CO VSF flattens, while the H$\alpha$ VSF steepens, and their amplitudes converge again at larger separations where pair statistics become insufficient.
    
            Despite the overall velocity alignment, the velocity dispersions differ: H$\alpha$ profiles are systematically broader than CO(2-1), with mean values of $(57 \pm 1)$~km~s$^{-1}$ versus $(26 \pm 2)$~km~s$^{-1}$, respectively. The average ratio is approximately 2:1, rising to 5:2 toward the edges -- though these extreme values may be affected by low SNR and should be treated with caution. Spaxel-by-spaxel comparison confirms this trend (Figure~\ref{fig:ALMA-vs-MEGARA_Maps-VSF}, panel A). While velocity centroids align closely, dispersion ratios range from 1:1 to 4:1, indicating consistently broader ionized gas line compared to molecular gas. This behavior is consistent with previous studies of multiphase filaments in BCGs (e.g., \citealt{Tremblay2018, Olivares2019}), which attribute broader H$\alpha$ lines to the higher volume filling factor of the warm ionized medium increasing the likelihood of multiple LOS velocity components. By comparison, the velocity dispersion measured for [\ion{C}{II}] in the southern filaments ($\sim 55$~km~s$^{-1}$ after instrumental resolution correction; \citealt{Werner2013}) aligns well with the H$\alpha$ values. While spectral resolution differences between instruments may partly contribute to this discrepancy, comparable velocity dispersions for the warm filaments are reported with similar resolution data \citep{Vigneron2024}. Finally, as suggested by \citet{Simionescu2018}, the low velocity dispersion of CO may reflect its embedding in a predominantly laminar flow.
            
        \subsubsection{Across the Central Filaments}
        \label{sec:multiple_vsfs}
            While VSFs are often applied globally, recent studies highlight their ability to reveal localized differences. For example, \citet{Li2020} and \citet{Ganguly2023} have demonstrated that VSFs can also capture distinctions between inner and outer filament regions, revealing how the nature of turbulence may vary with distance from the galaxy center. Another factor influencing VSF measurement is the viewing geometry. \citet{Hu2022} explored how projection effects -- both in simulations and observations -- can lead to directional dependencies in VSFs. By analyzing VSFs across different viewing angles and azimuthal regions in the plane of the sky, they showed that the VSF slope can appear significantly steeper when the LOS is aligned with the outflow direction, and flatter when viewed perpendicularly. Temporal evolution is another factor: \citet{Fournier2025} found that the VSF slope evolves over time, steepening during periods of AGN activity, such as jet-driven cavity inflation.
    
\begin{figure*}
                    \centering
                    \includegraphics[width=0.92\textwidth]{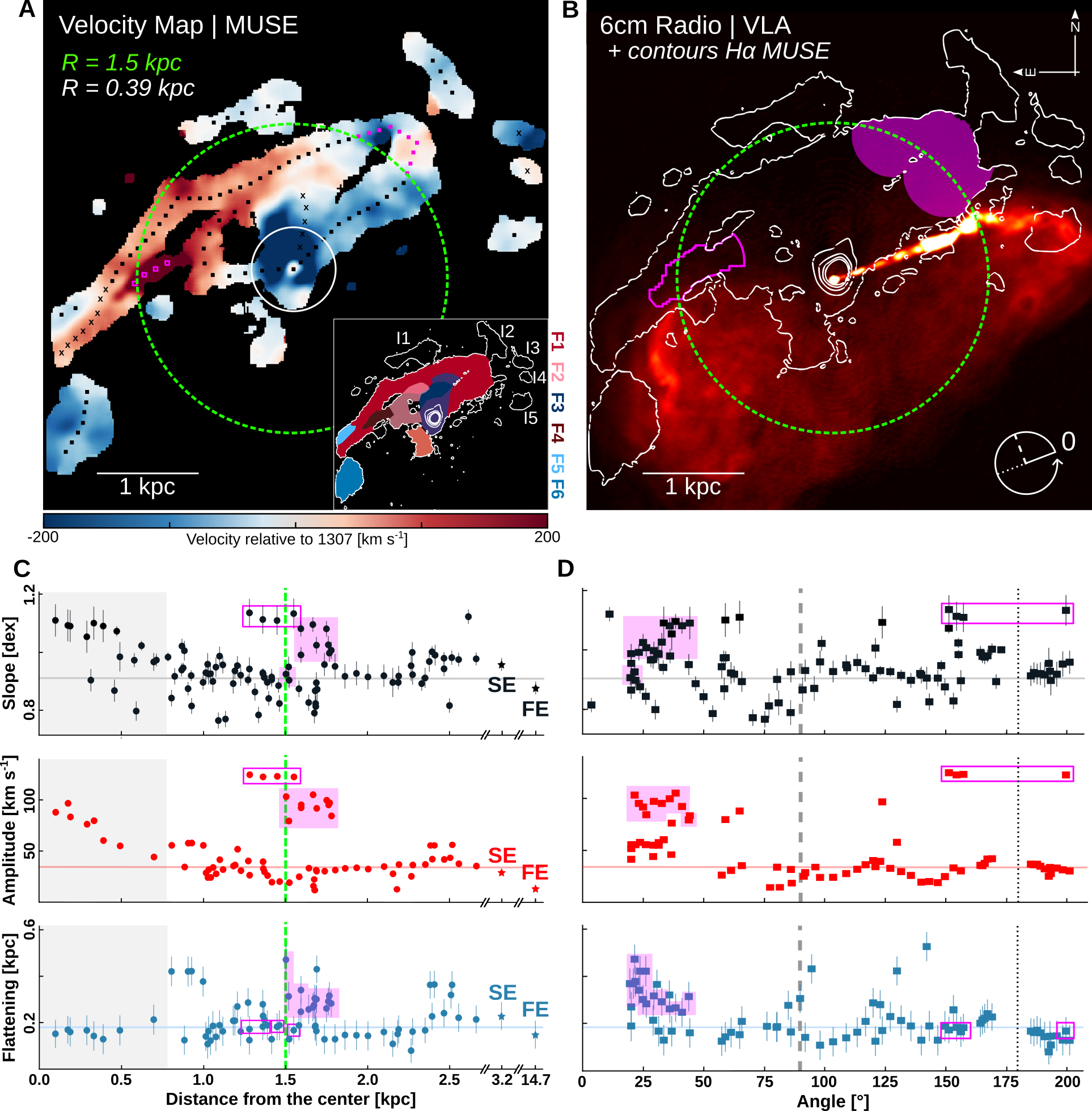}
                    \caption{\small (A) MUSE velocity map of the central filaments. Each marker corresponds to the center of a VSF computed within a circular aperture ($R =$ 0.39 kpc), shown as the white aperture, and distributed across subregions F1 to F5 and I1 to I5 (as defined in the reference map, lower right). Crosses indicate locations where no VSF flattening is observed. (B) VLA 6 cm radio emission with MUSE filaments contours overlaid in white. The schematic at bottom right shows the definition of the angle relative to the radio jet axis. Panels (C) and (D) present the VSF parameters -- slope (black), amplitude (red), flattening scale (blue) -- as functions of projected radial distance from the nucleus and projected angle with respect to the jet axis, respectively. Semi-transparent horizontal lines show the median values of each parameter: slope $\sim$0.92~dex, amplitude $\sim$32~km~s$^{-1}$, and flattening scale $\sim$0.18~kpc (excluding data points points within grey shaded and magenta-highlighted regions). In panel (C), a break in the horizontal axis separates measurements for the SE and FE filaments (star markers). The vertical dashed green line marks the projected length of the optical jet ($\sim$1.5 kpc), while the grey shaded region represents the 0.78~kpc diameter of the VSF aperture. In panel (D), vertical dashed and dotted lines correspond to angles of $90\degr$ and $180\degr$, respectively. Highlighted points in panels (A) through (D) are shown in magenta. Filled symbols indicate regions projected near the jet axis, while hollow symbols identify filament segments with velocities significantly offset from their local environments.}
                     \label{fig:VSFS_MUSE}
                \end{figure*}

            Given these known sources of variability in VSFs, and the localized signatures observed in the SE/FE filaments (Section~\ref{sec:SE-FE_VSFs}), we are motivated to investigate whether small-scale variations are present across other regions of the filamentary network. For this purpose, we use the MUSE velocity map (see Figure~\ref{fig:VSFS_MUSE}, panel A), which captures the central filaments and the area through which most of the visible jet / counter-jet extends. This dataset provides both the spatial coverage and a sufficient number of spaxels needed to compute localized VSFs across multiple subregions, enabling us to probe how turbulence varies locally with galactocentric distance and its relation to physical features such as the jet / counter-jet.
            
            Ideally, studying localized turbulence would require isolating individual filaments, but this is challenging due to projection effects. To reduce contamination from overlapping structures, we identified six filamentary subregions based on their projected morphology and kinematic continuity in the MUSE velocity map. This segmentation aims to separate structures that may be distinct in three dimensions yet appear close in projection and exhibit different velocities (see Appendix~\ref{appendix:vsf_subregions} for a comparison of VSFs with and without this separation). The segmentation is inherently subjective, relying on visual identification of morphological and kinematic continuity in projection. Ambiguity is unavoidable -- particularly at junctions where filaments overlap or intersect -- making it difficult to define strict boundaries. These uncertainties should be kept in mind when interpreting the resulting VSFs. In the following, we describe the subregions in order of increasing projected distance from the galaxy center, following the nomenclature of \citep{Sparks2004}, without implying any specific infalling or outflowing motion~: 
            
                \begin{description}
                    \item \textit{F1.} A prominent filament extending from the galaxy center northwestward before curving east. The 
                                       structure then forms a knot \citep{Sparks2004}, from which it diverges into a double-stranded morphology. The portion of the knot sharing similar velocities with the previous part of the filaments has been included in this subregion, as well as the southern strand that continues at an angle of $\sim 230\degr$ -- aligned with the direction of the previous straight segment. This strand gradually fades beyond the MUSE FOV. Overall, this subregion constitutes the longest coherent structures in projection.
                    \item \textit{F2.} A second filament emerging from the galaxy center and bending northeastward, forming a loop-like 
                                       structure that appears to rejoin F1 just north of the nucleus in projection.
                    \item \textit{F3.} A northern feature extending vertically from the center, composed of few narrow strands, which is 
                                       more clearly seen in [\ion{N}{II}] + H$\alpha$ images with the \textit{Hubble Space Telescope} (HST - \citealt{Werner2010, Tamhane2025}) and the \textit{ESO New Technology Telescope} \citep{Macchetto1996, Sparks2004}.
                    \item \textit{F4.} A small filament that connects the eastern end of F2 with F1, close to the knot. This structure 
                                       shows a distinct kinematic signature, with a velocity of $\sim 300$~km~s$^{-1}$, offset from neighboring filaments ($\sim$100~km~s$^{-1}$).
                    \item \textit{F5 \& F6.} Located at the southeastern edge of the MUSE FOV. F6 is associated with the SE filaments. F5 
                                        appears kinematically distinct from F1, with a velocity of $\sim -30$~km~s$^{-1}$ compared to $\sim 80$~km~s$^{-1}$ in adjacent regions. F5 traces the easternmost strand of the double-stranded morphology emerging after the knot described in F1, and gradually fades outside the MUSE FOV.
                \end{description}
            
             We also identified five small, spatially isolated filaments in the northern and western parts of the velocity map, labelled I1 to I5, named in clockwise order from the north. 
    
             Building on this segmentation, we defined a series of central positions along the isolated structures, spaced by $\sim100$~pc. Around each center, a localized VSF was computed from spaxels within a $\sim$0.39 kpc radius aperture -- chosen to match the widest filament widths in projection and to align with the MEGARA FOV. The results are not sensitive to the choice of radius, although changing it may shift the dominant turbulence scales being probed (see Sections \ref{sec:discussion_vsf} and \ref{sec:limitations_vsf}). From each VSF, we extract three key parameters to quantify turbulence: (1) the slope (power-law index), measured from the seeing limit up to the flattening -- or until velocity-pair statistics become insufficient if no flattening occurs; (2) the flattening scale, defined as the first point where the VSF flattens, which may indicate the characteristic injection scale of turbulent energy: (3) the amplitude, measured as the VSF velocity value at the flattening, providing an estimate of largest velocity associated with the dominant turbulent motions. Flattening scale and amplitude are only measured when flattening is present. The same parameters were also extracted from MEGARA VSFs for the SE (1 comp.) and FE filaments to directly compare central filaments from MUSE with those at larger projected distances. The full set of localized VSF regions and parameter variations -- slope, flattening scale and amplitude -- as a function of distance from the galaxy center ($\alpha$:~12h30m49.4233s, $\delta$:~+12d23m28.043s; NED) and relative angle to the visible jet (oriented at 21.6$\degr$, measured from the projected jet axis in the radio image) are shown in Figure~\ref{fig:VSFS_MUSE}. While substructures are defined to reduce projection contamination, they are not analyzed individually; VSFs show broadly similar properties across those substructures, as seen in the following paragraph (with exception of F4).

             The extracted VSF parameters exhibit spatial trends across the filament network. In particular, both slope and amplitude show a decline within the central $\sim0.8$~kpc and remain generally lower at larger projected distances and angular separation from the jet axis. In the center, slopes reach up to $\sim1.1$~dex decreasing below 1~dex beyond the aperture diameter. Similarly, VSF amplitudes peak at $\sim$100~km~s$^{-1}$ near the nucleus and fall below $\sim$50~km~s$^{-1}$ further out. Two groups of localized outliers are highlighted in magenta: one projected near the jet axis (filled) and one associated with the subregion F4 (unfilled). Both exhibit elevated slope and amplitude values; the filled outlier matches central measurements, while the unfilled one is slightly higher. In contrast, the flattening scales show weaker trends with galactocentric distance or jet orientation. Most are below 0.2~kpc, though a few reach $\sim$0.5~kpc, including the marked region near the jet axis. At larger distances, SE and FE filaments broadly follow the outer radial trend, with the FE filament showing slightly lower values. With respect to jet orientation, trends in VSF parameters are less pronounced. A modest decline from 0$\degr$ to 90$\degr$ may be present, but the highest values correspond to the innermost regions near the nucleus. The two outlier groups show distinct behaviors: the filled outlier close to the jet axis could reflect a genuine angular dependence, while the unfilled outlier deviates from other regions at similar position angles, likely due to local limitations (see Section~\ref{sec:discussion_vsf}). Overall, variations with angle appear to be shaped more by localized conditions than by a systematic geometric trend. Some southeastern filaments may also lie on the far side of the system, along the counter-jet direction not visible in observations \citep{Sparks1993}.
            
        \subsubsection{Interpretation of Localized VSFs}
        \label{sec:discussion_vsf}
            Localized VSFs across the M87 filament network reveal both coherent trends and notable deviations. A particularly sharp decrease in VSF slope and amplitude occurs within $\sim$0.8~kpc of the galaxy center. These enhanced values are primarily associated with regions where the VSF aperture overlaps the nucleus, where complex velocity fields arise from the superposition of filamentary structures and the presence of a nuclear rotating ionized disk (see \citetalias{Osorno2023}). In these cases, the resulting turbulence parameters are likely inflated due to superposition of distinct velocity components within a single aperture, rather than reflecting a purely physical change in the turbulence regime.
            
            Beyond the central region -- once the nuclear structure is excluded -- the VSF parameters appear to stabilize. As seen from the median lines in the parameter plots, the slope, amplitude, and the flattening scale remain relatively consistent across most filamentary substructures, particularly in the central and SE filaments (single-component), and to a lesser degree in the FE filament. This uniformity suggests a common small-scale turbulence feature, possibly linked to a shared driver. A first candidate is energy injection from Type Ia supernovae (SNe Ia), which occur in the old stellar populations of massive ellipticals. Simulations \citep{LiM2020, Mohapatra2024} show that SNe Ia drive compressive turbulence on $\sim$100-200~pc scales with velocity dispersions up to $\sim$40km~s$^{-1}$ -- consistent with the measured flattening scales and amplitudes. However, the greater distance of the FE filament from the main stellar body may call this explanation into question, though its slightly lower VSF parameters suggest they might not trace the same physical conditions. For this filament, this may reflect lower level of turbulence or structure formation, possibly linked to its location inside an older radio lobe associated with a past episode of AGN activity. In this scenario, the FE filament may be less actively driven, similar to the behavior on larger scales observed in outer filaments of many other galaxy clusters \citep{Li2020, Ganguly2023}. 
            
            Alternative explanations for the observed flattening could also naturally account for the steeper slopes seen in several VSFs. Similar steep slopes, exceeding the classical Kolmogorov expectation, have been reported in other systems (e.g., \citealt{Li2020, Ganguly2023}) and may arise when the VSF captures the influence of the primary driver (e.g., AGN feedback or SNe Ia) rather than the fully developed turbulent cascade. A possible explanation involves the partial dissipation of compressive modes on large scales prior to the transition to more purely Alfvénic turbulence on small scales with a Kolmogorov scaling, similar to what is observed in the solar wind plasma \citep{Hu2022}. Stratification may also affect the cascade across scales \citep{Mohapatra2022, Wang2023}, although its influence at the scales probed here is likely modest. Other plausible contributors to the observed deviations include magnetic tension \citep{Bambic2018, Grete2021}, plasma microinstabilities \citep{Kunz2014, Squire2020, Arzamasskiy2023}, gravitational acceleration \citep{Wang2021}, or ‘frozen-in’ supersonic turbulence \citep{Hu2022}. Altogether, these results suggest that local VSFs are highly sensitive to microphysical conditions and may reflect a superposition of regimes shaped by energy injection, dissipation processes and magnetic fields. 
            
            Another factor affecting the interpretation of flattening features is the finite size of the analysis region. In both SE and FE filaments, flattening often appears below the half-height of the analysis window (aperture), but the measured flattening scale itself tends to correspond closely to half the width of the filament enclosed within the window. This suggests the flattening may partly reflect a geometric distortion related to the filament’s cross-section within the aperture, rather than a genuine physical dissipation scale. However, tests of the window size (see Section~\ref{sec:limitations_vsf} and Appendix~\ref{appendix:vsf_fov}) show that the flattening persists at similar scales even when the aperture is increased and include filaments of varying, mostly larger, widths. 

            Among the localized VSFs, a few outliers stand out with notably elevated slope and amplitude values (highlighted in magenta in Figure~\ref{fig:VSFS_MUSE}). One of these lies near subregion F4, where the aperture covers overlapping filaments along the LOS -- leading to artificially inflate turbulence measures due to projection confusion. VSFs marked with crosses in Figure~\ref{fig:VSFS_MUSE} (panel A), showing no flattening, are also likely affected by the blending of kinematically distinct filaments (e.g., F2 and F5 near F4) or by nuclear contamination in F3. The second outlier lies northeast of the nucleus, near the jet axis and within a filament bend. This region shows a high LOS velocity ($\sim -275$~km~s$^{-1}$) and broad velocity dispersions (> 250~km~s$^{-1}$ in the map from \citetalias{Boselli2019}), in contrast to neighboring areas, where velocities typically range from $-50$ to $-100$~km~s$^{-1}$ and dispersion remain below 150~km~s$^{-1}$. These features may indicate genuine jet interaction or a geometrically thick, possibly LOS-elongated filament segment. Such a three-dimensional extension would increase LOS velocity spread, thereby boosting the VSF amplitude relative to adjacent regions. In contrast, the SE filament -- where distinct velocity components are more clearly resolved -- shows a VSF flattening only when both components are blended in a single fit (1 comp.), whereas no flattening appears in the VSF when computed using only the first component. This suggests that some observed flattenings elsewhere may arise from unresolved LOS superposition rather than true physical injection or dissipation scales, underscoring the importance of disentangling kinematic substructure in multiphase turbulence analysis.
            
            Finally, additional insight comes from comparing VSFs of the ionized and molecular gas in the SE filaments. Within the CO(2-1) detection region, both H$\alpha$- and CO-based VSFs exhibit a similar power-law slope at smaller scales. This behavior aligns with multiphase turbulence models, where cold, dense gas inherits the motions \citep{Gaspari2018, Gaspari2017a, Voit2018, Mohapatra2022}, and may also reflect direct coupling or dynamical interplay between phases at these scales. However, at larger separations (> $\sim$0.15~kpc), the VSFs diverge: the CO VSF steepens then flattens near $\sim$0.3~kpc, while H$\alpha$ maintains a more consistent slope. Notably, when the H$\alpha$ VSF is computed only within the CO detection footprint, it remains roughly parallel to the full SE-filament VSF but with a slightly lower amplitude. The divergence between the CO and H$\alpha$ VSFs within the CO-detected region may arise from a combination of factors -- including the differences in local volume-filling factor and projection effects. Additionally, the flattening of the CO VSF occurs near a scale comparable to half the width of the CO-emitting region ($\sim$300~pc east–west), suggesting that the finite spatial extent may also influence the VSF behavior (see Section\ref{sec:limitations_vsf}). Furthermore, to test whether the H$\alpha$ within the CO detection VSF behavior depends on aperture placement, we also computed it within a shifted CO-sized region elsewhere in the SE filaments. The small-scale slope is likely a robust feature, remaining consistent across apertures, while at larger scales (around the 1/2 window), the VSF steepens or flattens depending on location -- suggesting greater sensitivity to geometry and local structure.

        \subsubsection{Turbulent Heating in the FE filament}
        \label{sec:heating_rate}
            The dissipation of AGN-driven turbulence may serve as an important heating mechanism in cool-core clusters. In particular, \citet{Zhuravleva2014} showed that in M87 and Perseus, radiative cooling might be balanced by turbulent dissipation inferred from X-ray surface brightness fluctuations on large scales. However, X-ray observations lack the spatial resolution to probe turbulence on the smaller scales where most dissipation is expected to occur. \citet{Ganguly2023} extended this analysis to smaller scales by using H$\alpha$ filaments as tracers of turbulence in the central regions of a few systems, including M87 with MUSE WFM data. Here, we extend their approach to larger radii by analyzing the FE filament of M87 using MEGARA data.

            We estimate the turbulent heating rate, $Q_{\text{turb}}$, at the seeing scale (1\arcsec $\approx$ 78~pc), following the formalism of \citet{Ganguly2023} for VSFs steeper than the Kolmogorov slope
                \begin{equation}
                    Q_\text{turb} \approx 5.2\frac{\rho v_\text{l, 1D}^3}{l} \;,
                \end{equation}

            \noindent where $v_{l,\text{1D}}$ is the LOS velocity at scale $l$ from the VSF, and $\rho$ is the gas mass density given by $\rho = \mu m_p(n_i + n_e) = \mu m_p\xi n_e$. Here, $\mu = 0.61$ is the mean molecular weight, $m_p$ the proton mass, $\xi = 1.912$ accounts for total particle number (assuming 0.5 solar metallicity) and $n_e$ is the electron number density $n_e$. We compare this to the radiative cooling rate
                \begin{equation}
                    Q_\text{cool} = n_e^2(\xi - 1)\Lambda(T) \;,
                \end{equation}

            \noindent where $\Lambda(T)$ is the normalized cooling function computed from \citet{Schure2009}, assuming solar metallicity. At ICM temperatures ($10^7-10^8$~K), the metallicity dependence is weak and can be reasonably neglected.
            
            Adopting $n_e \approx 0.015$~cm$^{-3}$ and $T \approx 1.9$~keV at 15~kpc \citep{Zhuravleva2014}, we estimate a turbulent heating rate of $Q_\text{turb}\approx4.9~\times~10^{-28}$~erg~s$^{-1}$~cm$^{-3}$ and a radiative cooling rate of $Q_\text{cool}\approx~4.1~\times 10^{-27}$~erg~s$^{-1}$~cm$^{-3}$ for the FE filament. These results are consistent with findings from the central regions of M87 and other systems analyzed by \citet{Ganguly2023}, as well as with the behavior observed in numerical simulations (e.g., \citealt{Reynolds2015, Yang2016, Li2017}), where turbulent heating estimated on small scales typically accounts for only a minor fraction of radiative cooling (here, $\sim 10$\%). This contrasts with earlier X-ray–based estimates (e.g., \citealt{Zhuravleva2014, Zhuravleva2018}) that probe larger scales and may overestimate the contribution of turbulent dissipation. A more detailed discussion of these differences and their implications can be found in \citet{Ganguly2023}. We note that our estimated turbulent heating rate relies on the assumption that the kinematics of cool filaments and hot plasma are perfected coupled on small scales, which cannot be verified with existing X-ray telescopes yet due to their limited spatial resolution.

        \subsubsection{Limitations}
        \label{sec:limitations_vsf} 
            It is important to note that the VSFs are subject to several limitations and sources of uncertainty that may affect the interpretation of results at various level. Below, we review key limitations based on conclusions drawn from previous studies and a few tests conducted on our data. 

            \begin{enumerate}
                \item \textit{Projection Effects.} Projection effects along the LOS introduce two primary biases in the derivation of VSFs. 
                      (1) LOS integration can blend uncorrelated velocity components, artificially lowering $\langle \Delta v \rangle$ at smaller scales. By contrast, side-by-side filaments retain their velocity contrasts in projection, boosting $\langle \Delta v \rangle$ at a given scale. Both effects are amplified in localized VSFs, where small apertures and limited spaxel coverage increase the likelihood of overlapping or adjacent structures dominating the measurement. This can steepen slopes and inflate amplitudes, particularly in regions like F4, where adjacent structures differ significantly in velocity (see Section~\ref{sec:multiple_vsfs}). (2) Projection compresses true 3D separations into smaller 2D distances, shifting large velocity differences to artificially small scales and flattens the observed VSF. Numerical simulations \citep{ZuHone2016, Mohapatra2022, Fournier2025} demonstrate that projection steepens VSFs at small scales due to LOS cancellation but flattens them at larger scales where uncorrelated motions dominate. Although these competing effects cannot be disentangled without full 3D spatial information, some insight into LOS complexity can be gained from the spectral structure itself. In particular, multi-component line profiles offer observational evidence for unresolved kinematic substructure along the LOS -- as seen in other systems (e.g., \citealt{Hatch2006, Canning2011, Hamer2016, Vigneron2024}) -- though their influence on VSFs has only recently been studied. \citet{Zhang2022} and \citet{Li2025} show that in compact sources, projection can flatten VSFs by introducing high-$|\delta v|$ pixel pairs at small separations. Our results in the SE filaments (Section~\ref{sec:SE-FE_VSFs}) support this: approximately 23\% of spaxels exhibit multiple velocity components, and the $\sim$0.3~kpc flattening vanishes when the VSF is computed using only the first component. Moreover, the slope becomes slightly steeper in the single-component case, reinforcing that LOS superposition can mask true velocity correlations and bias turbulence estimates. We note that while MEGARA's resolution ($R \sim 6000$) allows us to resolve multiple components, finer substructure may remain unresolved. Higher-resolution data could better constrain VSFs and reveal additional complexity. 

                \item \textit{Seeing Effects.} Atmospheric seeing impacts VSFs, particularly at small scales.
                      Using MUSE data of M87, \citet{Li2020} applied Gaussian smoothing matched to the point-spread function (PSF) and found a modest ($<$0.1~dex) suppression of VSF power due to velocity homogenization between neighboring spaxels. Similar studies \citep{Chen2023, Ganguly2023} report that smoothing steepens the VSF slopes below the PSF scale while preserving the main flattening. Our MEGARA-based tests using the same approach show similar suppression -- despite fewer velocity pairs due to narrower filaments. Simulations by \citet{Fournier2025}, modeling AGN-driven turbulence with the XMAGNET framework, further confirm that smoothing artificially steepens VSF slopes below the kernel scale. Therefore, we consider the VSFs unreliable below the seeing limit. 

                \item \textit{Spaxel size \& Binning.} Our datasets with span different spaxel sizes, from 0.2$\arcsec$ (MUSE) 
                      and 0.31$\arcsec$ (MEGARA) to 0.64$\arcsec$ for SITELLE (after 2~$\times$~2 binning), raising concerns about their effect on VSFs. To evaluate this, we conducted a series of spaxel size/binning tests using SITELLE, MEGARA and MUSE data. As detailed in Appendix~\ref{appendix:vsf_spaxelsize}, we compared matched subregions using the original SITELLE spaxel size and both original and resampled MUSE data. We also applied various binning schemes to MUSE (up to 10~$\times$~10) and MEGARA (3~$\times$~3). In all cases, the overall VSF shapes remain relatively stable. Larger spaxel sizes occasionally steepen the slope at small scales, but flattening scale and amplitude stay consistent. These results align with simulations (e.g., \citealt{Fournier2025}), which report that resolution degradation can bias small-scale VSF properties. Although such effects are relatively minor within our tested spaxel range, they should nonetheless be kept in mind when interpreting VSFs. Notably, SITELLE consistently shows higher amplitudes than MUSE or MEGARA, even after matching spaxel size. We therefore attribute this difference to the noise of SITELLE's velocity map, which introduces larger spaxel-to-spaxel variations and enhances apparent velocity differences.

                \item \textit{Window Size.} To evaluate whether the size and geometry of the analysis window affects VSF measurements, we used the full MUSE velocity map of the central filaments and computed VSFs within circular and rectangular windows of varying sizes. While only the circular-window results are shown in Appendix~\ref{appendix:vsf_fov}, we find that both geometries yield consistent VSF shapes. Across window sizes, the first flattening scale and amplitude remain stable, suggesting that small- and intermediate-scale dynamics are unaffected. Differences mainly appear at larger separations, where a second flattening or a diminished first flattening can emerge due to contributions from large-scale dynamics. This implies that window size does not strongly bias VSFs within the explored FOV range. However, larger windows naturally incorporate broader scales, potentially shifting the dominant dynamics from local turbulence to global kinematics. Few studies (e.g., \citealt{Ha2022, Li2025}) have tested the impact of window size on VSFs, showing that beyond separations of roughly half the map size, edge effects and reduced pair statistics may lead to artificial flattenings. Although our own test yields broadly consistent trends, this underscores the need for caution when interpreting VSFs at the largest scales.

                \item \textit{Noise.} The impact of noise on VSFs has been explored in previous work. \citet{LiM2020} demonstrated that 
                      higher noise levels, modelled through lower SNR cuts, can raise the VSF amplitude at small scales and slightly alter the slope, while leaving the flattening unaffected. This may partly explain the systematically higher VSF amplitudes seen in SITELLE, likely driven by increased noise and its lower spectral resolution.
            \end{enumerate}

\section{Ionization Sources}
\label{sec:ionization_sources}

    \subsection{Line Ratios}
    \label{sec:line_ratios}
        Several emission lines in the MEGARA and SITELLE data -- specifically [\ion{O}{I}]$\lambda$6300, H$\alpha$, [\ion{N}{II}]$\lambda$6583 and [\ion{S}{II}]$\lambda$6716,~6731 -- provide diagnostics of the physical conditions of the ionized gas. As a first step, we examine key emission-line ratios sensitive to ionization mechanisms such as shocks: [\ion{O}{I}]/H$\alpha$\footnote{[\ion{O}{I}]$\lambda$6300.}, [\ion{N}{II}]/H$\alpha$ and [\ion{S}{II}]/H$\alpha$\footnote{([\ion{S}{II}]$\lambda$6716 + [\ion{S}{II}]$\lambda$6731])/H$\alpha$.}.
        
        Due to the faintness and low SNR of the [\ion{O}{I}]$\lambda$6300 line in MEGARA spaxels, spatially resolved [\ion{O}{I}]/H$\alpha$ maps could not be constructed. Instead, we measure the [\ion{O}{I}]/H$\alpha$ ratio from the integrated MEGARA spectra of the SE and FE filaments, obtaining low values of 0.23 and 0.16, respectively. These results are consistent with previous measurements based on MUSE WFM data for the SE filaments ($\sim$0.2; \citetalias{Boselli2019}) and long-slit spectroscopy from the \textit{Observatoire de Haute Provence} (OHP) for the FE filament (0.18; \citetalias{Gavazzi2000}).
        
\begin{figure*}
                \centering
                \includegraphics[width=0.85\textwidth]{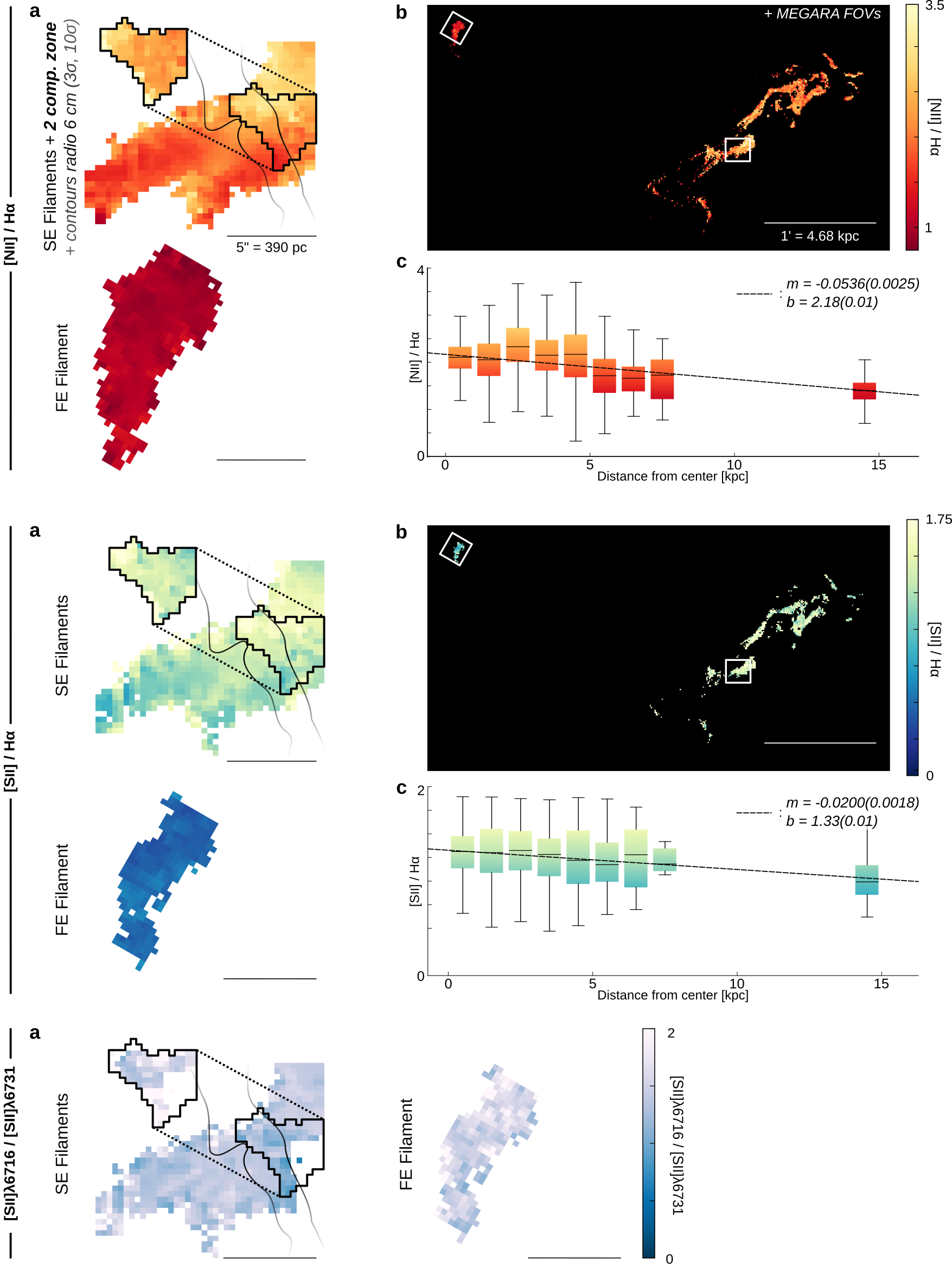}
                \caption{\small Maps of the [\ion{N}{II}]/H$\alpha$ (top panel), [\ion{S}{II}]/H$\alpha$ (middle panel) and [\ion{S}{II}]$\lambda$6716/[\ion{S}{II}]$\lambda$6731 (bottom panel) line ratios, using only spaxels with SNR$_{\text{H}\alpha} > 3$ and SNR$_{\text{[SII]}\lambda6716} > 3$ where applicable, derived from (a) MEGARA and (b) SITELLE observations. In the MEGARA maps, black gradient contours over the SE filament indicate the 6~cm VLA radio emission tracing the counter-jet lobe at 3$\sigma$ and 10$\sigma$ \citep{Hines1989}. In the SITELLE maps, the MEGARA pointings targeting the SE and FE filaments are outlined in white. (c) A radial gradient was also derived for the two first line ratios with SITELLE data, using concentric annuli centered on the galaxy nucleus. Distributions of the line ratio within each annulus are shown as boxplots, from which the median values were used to perform a linear regression. The best-fit parameters of the regression are reported in the upper right corner of the panel. Color scales are consistent across maps of the same type. All scale bars correspond to 5$\arcsec$.}
                 \label{fig:Ratio_Maps-Gradients}
            \end{figure*}

        The [\ion{N}{II}]/H$\alpha$ and [\ion{S}{II}]/H$\alpha$ ratios are sufficiently strong to be mapped with SITELLE and MEGARA. Figure~\ref{fig:Ratio_Maps-Gradients} presents these ratio maps along with radial gradients derived from SITELLE data, computed using concentric 1~kpc-wide annuli centered on the galaxy nucleus. Tests confirmed that the analysis is insensitive to annulus spacing. Within each annulus, line ratio distributions are summarized with boxplots (median, interquartile range, and overall spread), and a weighted linear regression on median values to quantify the radial trends. Overall, SITELLE results are consistent with MEGARA and MUSE WFM but systematically slightly higher, likely due to increased noise levels. In the central regions, SITELLE shows high [\ion{N}{II}]/H$\alpha$ ratios of $\sim$2. Beyond $\sim$2.5~kpc, the gradient does not decline monotonically: local variations are observed between 2.5–5~kpc, especially toward the southern filament. Farther out, the ratio gradually decreases to $\sim$1.5 at $\sim$7.5~kpc. A similar, though shallower, trend is seen in the [\ion{S}{II}]/H$\alpha$ profile, decreasing from $\sim$1.3 to $\sim$1.1, with greater dispersion at intermediate radii. MEGARA observations of the SE filaments confirm the elevated ratios seen with SITELLE. The [\ion{N}{II}]/H$\alpha$ ratio decreases from $\sim$2.75 near the nucleus to $\sim$1.95 further out, with the highest values projected along the edge of the radio lobe. This regions also corresponds to where two velocity components are detected, both exhibiting similar elevated values around $\sim$2.5. The [\ion{S}{II}]/H$\alpha$ ratio follows a similar pattern, declining from $\sim$1.51 to $\sim$1.24, with both components near $\sim$1.45. \citet{Werner2013} and \citetalias{Boselli2019} similarly reported high [\ion{O}{III}]/H$\beta$ ratios in this same region ($\sim$2) -- exceeding those measured in the galaxy nucleus -- consistent with Seyfert-like ionization. By contrast, the FE filament displays significantly lower ratios in both datasets. The [\ion{N}{II}]/H$\alpha$ ratio remains near $\sim$1.0, and the [\ion{S}{II}]/H$\alpha$ ratio near $\sim$0.5, consistent with a lower ionization environment. Although we do not measure the [\ion{O}{iii}] and H$\beta$ lines in our new data, we recall that \citet{Gavazzi2000} reported a [\ion{O}{iii}]/H$\beta$ ratio of $\sim$0.85 in the same region.  
        
        [\ion{S}{II}]$\lambda$6716/[\ion{S}{II}]$\lambda$6731 ratio maps were also produced from the MEGARA data (see Fig.~\ref{fig:Ratio_Maps-Gradients}), serving as a tracer of the electron density $n_e$ of the ionized gas \citep{Osterbrock2006}. This analysis was not performed with the SITELLE data due to its lower spectral resolution and insufficient SNR for [\ion{S}{II}]$\lambda$6731 across the field. In the SE filaments, the mean [\ion{S}{II}] doublet ratio is $\sim1.37$, corresponding to $n_e \sim 40$ cm$^{-3}$ assuming a electron temperature of 10,000~K and using the calibration from \citet{Proxauf2014}. This ratio is consistent with previous measurements, including $\sim 1.40$ reported by \citet{Werner2013} and \citetalias{Boselli2019} at the edges of the SE filaments. In the FE filament, we measure a higher mean ratio of $\sim 1.56$, which lies above the low density limit with $n_e \sim 10$ cm$^{-3}$, as reported by \citetalias{Gavazzi2000}. 
        
    \subsection{WHAN Diagram}
    \label{sec:whan_diagram}
        To further investigate the nature of the ionizing sources in the SE and FE filaments, we use the Width H$\alpha$-Nitrogen (WHAN) diagram \citep{CidFernandes2010, CidFernandes2011}, traditionally used to classify galaxies by dominant ionization mechanisms. It an alternative to the classical Baldwin, Phillips \& Terlevich (BPT) diagram \citep{Baldwin1981} when key emission lines such as [\ion{O}{III}]$\lambda$5007 and H$\beta$ are unavailable, as in our MEGARA and SITELLE data. However, these lines have been detected with other instruments in the central \citepalias{Boselli2019}, southern \citep{Werner2013} and in the outer FE filament \citepalias{Gavazzi2000}. The WHAN diagram combines the [\ion{N}{II}]/H$\alpha$ line ratio with the H$\alpha$ equivalent width $W_\alpha$, defined as line flux over local continuum flux, measuring line strength relative to the stellar background.
        
        In this framework, we use well-established diagnostic lines to guide our interpretation. Specifically, regions with log([\ion{N}{II}]/H$\alpha$)~$>~0.4$ and $W_\alpha~>~6$~\si{\angstrom} are classified as Seyfert-like, indicative of strong ionizing sources such as AGN; those with 3~\si{\angstrom}~$<~W_\alpha~<~6$~\si{\angstrom} fall into the LINER-like regime, typically associated with weaker ionization. Values of $W_\alpha~<~3$~\si{\angstrom} correspond to retired or passive regions, where ionization is likely driven by old stellar populations. Star-forming regions are found at log([\ion{N}{II}]/H$\alpha$)~<~-0.4 and $W_\alpha~>~3$~\si{\angstrom}, consistent with photoionization by young massive stars. Here, we use the WHAN diagram for the SE and FE filaments -- not to impose a strict classification, but to identify regions of relatively higher or lower energy ionization source, acknowledging the likely contribution of multiple excitation mechanisms \citep{Ferland2009, McDonald2012, Rhea2025}. 
        
            \begin{figure*}
                \centering
                \includegraphics[width=\textwidth]{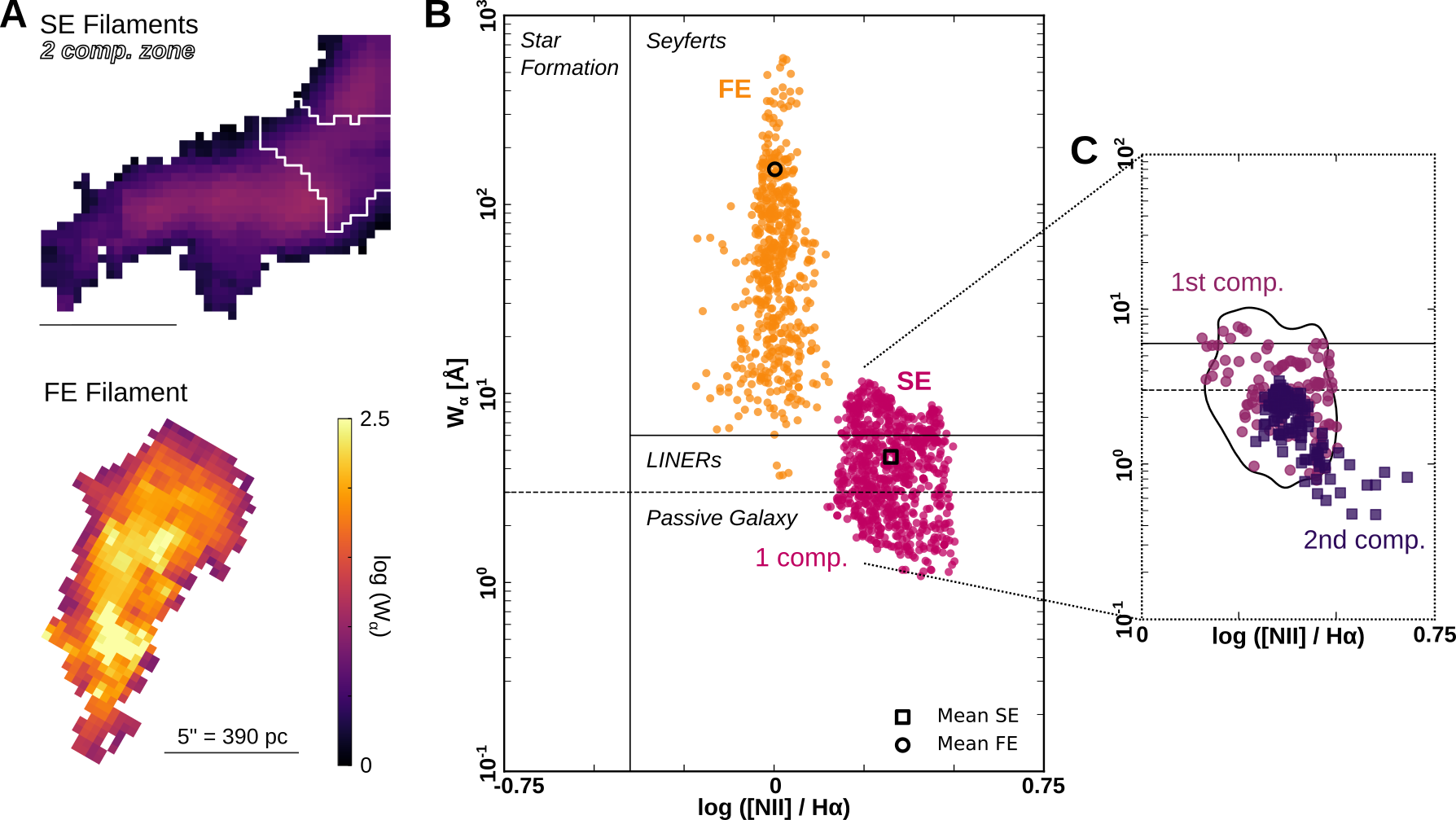}
                \caption{\small (A) Maps of the H$\alpha$ equivalent width $W_\alpha$ for the SE and FE filaments observed with MEGARA (SNR$_{\text{H}\alpha} > 3$). The white outline marks the two-component region within SE filaments. (B) WHAN diagrams constructed spaxel-by-spaxel with each filament (SE in pink~-~1~comp., FE in orange). The mean values for each distribution are marked by a black square (SE) and a black circle (FE). (C) WHAN diagram showing the first and second components within the SE two-component region. Both components largely overlap with the single-component SE distribution (black contour).}
                \label{fig:WHANDiagrams}
            \end{figure*} 

        Figure~\ref{fig:WHANDiagrams} presents the WHAN diagram on a spaxel-by-spaxel basis for the SE and FE filaments, along with the equivalent width maps. In the SE region, where two components were identified, both components are shown separately. No spaxels fall in the star-forming WHAN region, consistent with the high [\ion{N}{ii}]/H$\alpha$ ratios presented in Section~\ref{sec:line_ratios}. The FE filament primarily occupies the Seyfert-like region, with a mean $W_\alpha \sim 150$~\si{\angstrom}. In contrast, the SE filaments initially show much lower $W_\alpha$ values (mean $\sim$5\si{\angstrom}) due to dilution by the underlying stellar continuum, placing spaxels in the passive, LINER-like, or Seyfert-like regions.
        
        We note that both filaments exhibit broad spread in $W_\alpha$. Measurement uncertainties may contribute but geometry -- including projection effects, varying thickness, partial LOS overlap -- likely plays a role. To assess measurement impact, we examined $W_\alpha$ versus H$\alpha$ SNR, finding strong positive correlations (Pearson coefficients 0.93 for SE, 0.76 for FE). This sensitivity highlights limitations of spaxel-level WHAN diagnostics and suggests integrated-region analyses yield more reliable ionization classifications.

    \subsection{Velocity Dispersion vs Line Ratios}
    \label{sec:velocity_dispersion_line_ratios}
        Combining ionization-sensitive line ratios with kinematics can also offer insight into the excitation mechanisms. In particular, diagrams comparing line ratios to velocity dispersion are useful to identify shocks, which tend to broaden lines and enhance low-ionization transitions such as [\ion{N}{II}], [\ion{S}{II}], and [\ion{O}{I}] relative to H$\alpha$ \citep{Rich2011, Rich2015, McDonald2012, Ho2014, Boselli2019, Johnston2022}. In contrast, photoionized regions typically exhibit both lower line ratios and narrower velocity dispersions.

\begin{figure*}
                \centering
                \includegraphics[width=\textwidth]{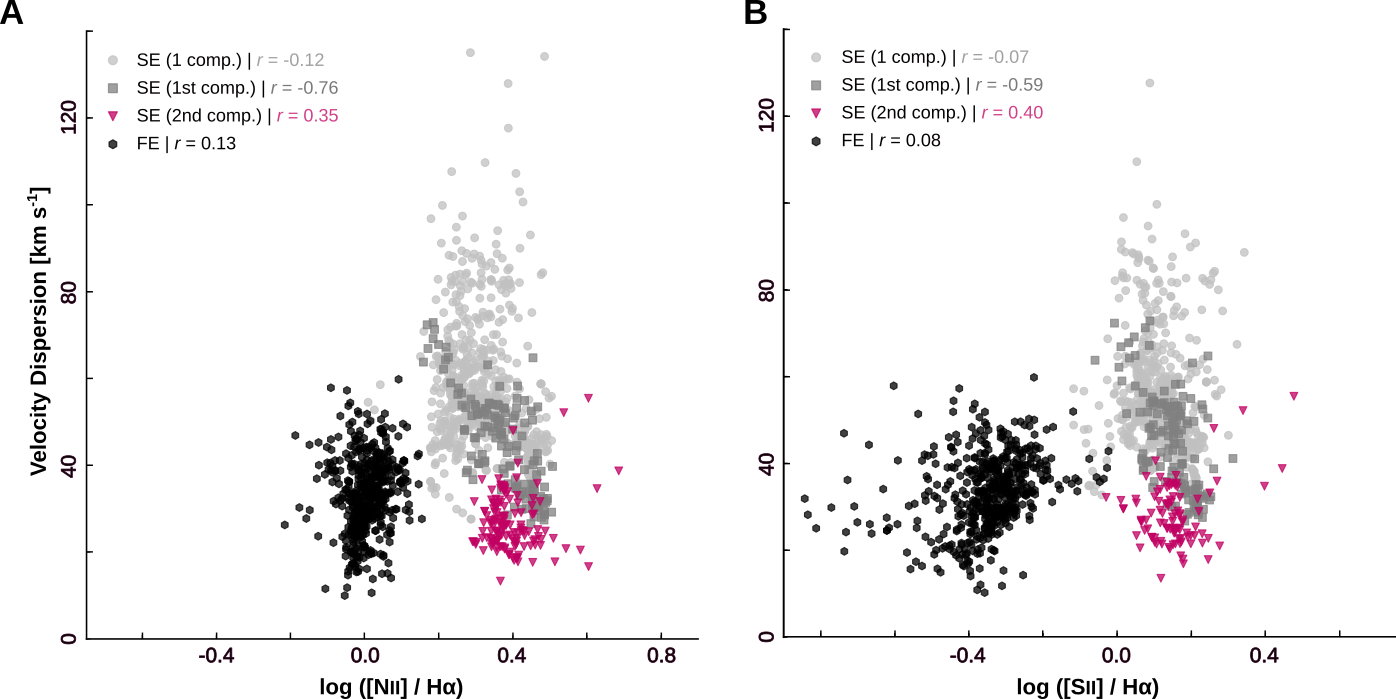}
                \caption{\small Spaxel-by-spaxel diagrams comparing the velocity dispersion to (A) [\ion{N}{II}]/H$\alpha$ and (B) [\ion{S}{II}]/H$\alpha$ for the SE and FE filaments from MEGARA data. Data points from the SE filaments are shown in light grey circles (1 comp.), dark grey squares (1st comp., in the two-component region) and pink triangle (2nd comp.). The FE filament is represented by black hexagons. The Pearson correlation coefficient $r$ is indicated in the upper-left corner or each panel; values near $\pm$1 indicate strong positive/negative correlation, while values near 0 indicate none.}
                 \label{fig:VelocityDispersion-LineRatios}
            \end{figure*}

        In M87, \citetalias{Boselli2019} identified a clear positive correlation between velocity dispersion and [\ion{O}{I}]/H$\alpha$ in the central filaments, consistent with AGN-driven shocks, with weaker or absent correlations for [\ion{N}{II}]/H$\alpha$ and [\ion{S}{II}]/H$\alpha$ at larger distances. Here, we extend this analysis to the SE and FE filaments using MEGARA data to assess whether similar trends persist farther from the nucleus. We constructed spaxel-by-spaxel diagrams of velocity dispersion versus [\ion{N}{II}]/H$\alpha$ and [\ion{S}{II}]/H$\alpha$ for each pointing and velocity component (where applicable), and quantified the trends using the Pearson correlation coefficient $r$. These results are presented in Figure~\ref{fig:VelocityDispersion-LineRatios}. 
        
        No strong or significant correlations emerge in either filament. [\ion{N}{II}]/H$\alpha$ generally yielded slightly higher $r$ values than [\ion{S}{II}]/H$\alpha$. Overall, both filaments show relatively high line ratios but low velocity dispersions -- typically <100~km~s$^{-1}$, and only a few tens of km~s$^{-1}$ in the FE filament, consistent with earlier results \citepalias{Gavazzi2000}.
    
    \subsection{Interpreting the Ionization Mechanisms}
    \label{sec:interpretion_ionization}
        The spatial and spectral properties of warm ionized gas in M87 reveal a complex excitation environment with multiple ionization processes contributing. Our results support previous findings that photoionization by young massive stars is unlikely to dominate, given the absence of compact UV or optical stellar counterparts associated with the filaments (\citealt{Werner2013}; \citetalias{Boselli2019}; \citealt{Tamhane2025}) and the elevated low-ionization line ratios observed throughout the system. The decoupling between the ionized gas and the stellar continuum, particularly in the FE filament, further rules out stellar sources and instead implies the presence of a harder excitation mechanism. Although, we note that post–asymptotic giant branch (post-AGB) stellar populations can produce LINER-like emission (e.g., \citealt{Lagos2022}). Such line ratios are primarily observed in the central region (as shown in \citetalias{Boselli2019}) and in the SE filaments.
        
        Ionization conditions vary spatially. Within the central $\sim$0.5~kpc, [\ion{N}{II}]/H$\alpha$ and [\ion{S}{II}]/H$\alpha$ ratios are high, coinciding with broad velocity dispersions and previously reported elevated [\ion{O}{I}]/H$\alpha$ and [\ion{O}{III}]/H$\beta$ line ratios from MUSE \citepalias{Boselli2019}. Collectively, these diagnostics point toward excitation dominated by the AGN.
        
        The SE filaments exhibit similarly high, or even higher, line ratios -- especially near the projected boundary of the counter-jet radio lobe. Ratios decline sharply beyond the region where two velocity components are detected, just outside the radio lobe' edge. This spatial correspondence suggests direct mechanical interaction between the AGN outflow and the filamentary gas, likely via shocks or compressive flows. However, we find no clear correlation between line ratios and velocity dispersion -- a classical signature of fast radiative shocks. Similar results have been reported in other BCGs (e.g., \citealt{Hamer2016, Hamer2019}), where high-excitation line ratios lack the broad profiles predicted by shock models. Moreover, the observed line ratios -- particularly [\ion{O}{I}]/H$\alpha$ and [\ion{S}{II}]/H$\alpha$ -- disagree with the predictions of fast shock models (e.g., \citealt{allen2008mappings}). The relatively low velocity dispersions measured here suggest either weak shocks or additional ionization sources. Plausible alternatives include collisional heating by cosmic rays, potentially energized by magnetohydrodynamic waves along the filaments \citep{Ferland2009, Fabian2011, McDonald2012}, dust scattering \citep{Mattila2007, Seon2015}, thermal conduction from the surrounding hot ICM, and ionization by soft X-ray/extreme ultraviolet (EUV) photons from the cooling ICM \citep{Polles2021}.

        In contrast, the FE filament has a lower ionization state, with significantly lower line ratios and narrow line widths. Nonetheless, the elevated H$\alpha$ equivalent widths, persistent [\ion{N}{ii}]/H$\alpha$ ratio, and lack of underlying stellar continuum suggest that stellar photoionization is not the dominant excitation mechanism. \citetalias{Gavazzi2000} interpreted this outer filament as relic of past AGN activity, potentially energized by intermediate-velocity shocks from the now-inactive eastern radio lobe -- a scenario consistent with its morphology, location, and ionization properties. Additionally, the [\ion{O}{I}]/H$\alpha$ ratio again disagrees with fast shock models \citep{allen2008mappings}, and the absence of broadened lines may point to fossil or decaying shocks, or steady-state processes such as cosmic ray heating, turbulent mixing layers operating at lower energies, or X-ray emitting gas from the cooling ICM (e.g., \citealt{Fabian2011, McDonald2012, Hamer2016, Polles2021}).

\section[Hα -- X-Ray Correlation]{H$\alpha$ -- X-Ray Correlation}
\label{sec:Ha-RX_correlation}
    A spatial correlation between H$\alpha$ + [\ion{N}{II}] emission and soft X-ray structures in M87 has long been noted, particularly within the central $\sim$5~kpc \citep{Young2002, Sparks2004, Werner2010, Werner2013}. These studies highlighted a close morphological association between warm ionized gas and the cooler X-ray-emitting phase ($\sim0.5$~keV), suggesting thermal coupling and/or co-evolution. More recently, \citet{Olivares2025} quantified a strong correlation between H$\alpha$ and X-ray surface brightness in seven cool-core clusters, including M87 using the MUSE WFM data covering the central $\sim1\arcmin~\times~1\arcmin$ region. In this work, we revisit this correlation using SITELLE's larger FOV, covering the southeastern and FE filaments beyond the central region. 

    We adopt the same methodology as presented in \citet{Olivares2025} to compare the H$\alpha$ and soft X-ray surface brightness in M87. We use deep archival \textit{Chandra} observations of M87, from which the filamentary X-ray component is isolated using the Poisson Generalized Morphological Component Analysis (pGMCA; \citealt{Bobin2020}). The pGMCA method can extract a user-defined number of components with no prior physical or instrumental knowledge, by focusing on the morphological diversity of the different emissions. We decompose the X-ray emission into four components reflecting a mix of structures: diffuse halo, AGN cavities, and filaments -- without a strict one-to-one association. In our case, the filamentary emission is captured across two distinct components. The first highlights the sharper, well-defined filaments, broadly consistent with the morphology of the warm ionized gas. The second component also traces some of these filamentary structures, but additionally includes more extended regions lacking a clear H$\alpha$ counterparts -- notably encompassing parts of the X-ray arms (e.g., southwestern arm). For the warm ionized phase, we use SITELLE H$\alpha$ map (binned $2 \times 2$). Filamentary regions are identified manually on this map, following coherent morphological structures in projection, and applied to the X-ray filament maps to ensure spatial correspondence. For each region, we extract both the H$\alpha$ and soft X-ray fluxes, compute the luminosities using the redshift-dependent luminosity distance, and derive the surface brightness (in erg s$^{-1}$ kpc$^{-2}$) by dividing by the physical area. Regions contaminated by the central AGN or background sources are excluded. Details of the surface brightness fitting and uncertainty treatment is provided in \citet{Olivares2025}.

        \begin{figure*}
            \centering
            \includegraphics[width=\textwidth]{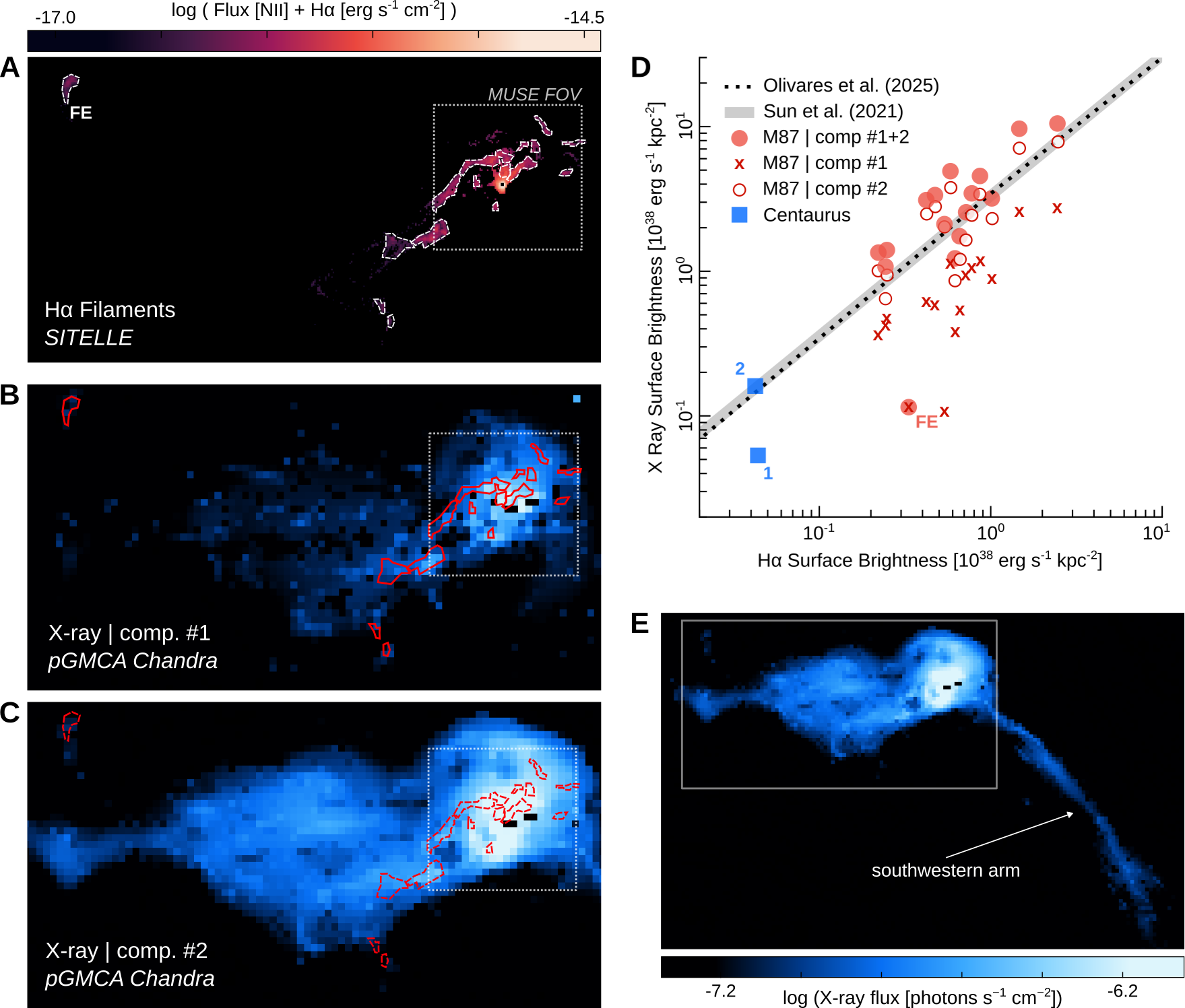}
            \caption{\small (A) H$\alpha$ flux map (SNR$_{\text{H}\alpha}$ > 3, 2~$\times$~2 binned). (B) X-ray flux map of the first component from the decomposition. (C) X-ray flux map of the second component. In panels A-C, surface brightness extraction regions are show in white/red (same regions across panels), and the MUSE WFM pointing is indicated by the light grey dotted box. (D) H$\alpha$-X ray surface brightness correlation for filaments in M87 (red, representing the sum of both X-ray components) and two isolated clumps in Centaurus (blue squares). For M87, crosses and white circles with red contours represent values derived using the first and second X-ray components, respectively. Errors are smaller than the scatter. The dashed black line shows the best-fit relation from \citet{Olivares2025}, and the thick grey line corresponds to the diffuse gas trend in stripped tails from \citet{Sun2022}. (E) Full FOV X-ray map of the second component, including the southwestern arm. The zoomed-in region shown in panels B and C is indicated by the box.}
            \label{fig:Ha-XR_correlation}
        \end{figure*} 

    Figure~\ref{fig:Ha-XR_correlation} shows the H$\alpha$ - X-ray surface brightness correlation across M87's filaments, using the two X-ray components identified as containing filamentary emission. The first and second components yield individual X-ray/H$\alpha$ surface brightness ratios of 1.3$^{+0.6}_{-1.1}$ and 3.7$^{+2.7}_{-2.3}$, respectively. Summed together, they reproduce the tight correlation previously observed with MUSE WFM \citep{Olivares2025}, now extend to the larger filament network captured by SITELLE. The typical X-ray/H$\alpha$ surface brightness ratio for the combined components (excluding the FE filament) is 5.0$^{+3.4}_{-3.0}$, consistent with the cool-core average of 4.1$\pm$2.4 found in that study. This consistency reinforces the idea that local excitation mechanisms (e.g., cosmic ray heating, turbulent mixing layers, reprocessed X-ray radiation) govern the emission, while the distance to the AGN plays little role. The persistence of a strong correlation across diverse environments, including stripped galaxy tails \citep{Sun2022}, supports a common mechanism linking hot and warm gas. While this behaviour is consistent with models in which local thermal instabilities in the ICM lead to condensation of warm gas \citep{McCourt2012, Sharma2012, Li2014}, it may more broadly reflect ongoing dynamical interactions between warm filaments and the surrounding hot atmosphere -- regardless of whether the warm gas formed via precipitation or uplift-driven processes. 
    
    Intriguingly, one region in M87 -- the isolated FE filament (labelled ‘FE’ in Fig.~\ref{fig:Ha-XR_correlation}) -- lies well below the main correlation with X-ray/H$\alpha$ ratio of 0.35. A similar deviation is found in Centaurus cluster, where we applied the same pGMCA-based analysis with four X-ray components. While \citet{Olivares2025} found that Centaurus' central filaments follow the main trend, our extended analysis identifies two more isolated northeastern clumps. Among these, the clump located at 7-9~kpc (clump 1 in panel D of Fig.~\ref{fig:Ha-XR_correlation}) presents a low X-ray/H$\alpha$ ratio of 1.2. In both Centaurus and M87, the deviating clumps lie near the inner edges of extended soft X-ray features likely shaped by past AGN activity: the northeastern Centaurus clump aligns with the direction of a ghost cavity \citep{Crawford2005, Sanders2016}, while M87's FE filament sits within an old radio lobe. This suggests that such clumps were uplifted during earlier episodes of radio-mode feedback and have since cooled  more, evolved differently or become less coupled to the surrounding hot ICM compared to the core-connected filaments. Alternatively, the observed discrepancy may reflect limitations in isolating diffuse filamentary X-ray emission using pGMCA: as the algorithm can extract components based on their morphological features, it will more easily retrieve sources that are extended with a characteristic shape than small clumps -- particularly in low surface brightness regions.

\section{Summary and conclusions}
\label{sec:conclusions}
    We presented a comprehensive analysis of the warm ionized filaments in M87, combining new and archival multiwavelength data to probe their kinematics, turbulence, ionization state, and multiphase connections in different environments shaped by AGN feedback. MEGARA observations target two key regions: the SE filaments, which host the only known molecular cloud detected with ALMA, and the isolated FE filament within an old radio lobe. SITELLE offers full coverage of the filamentary system, enabling a global view of the ionized gas. These datasets were complemented by MUSE WFM observations of the central filaments and \textit{Chandra} X-ray data tracing the surrounding hot plasma. Our main findings are as follow:

        \begin{enumerate}
            \item M87's ionized filaments exhibit complex motions that reflect strong local influences (e.g., jet interaction) beyond simple rotation. SITELLE maps the entire velocity field for the first time, reaching outer filaments skirting buoyant bubbles of different ages with velocities from $\sim -$250 to 50~km~s$^{-1}$. MEGARA resolves a two-component structure in the SE filaments near the radio lobe edge, separated by $\sim$70~km~s$^{-1}$, with similar velocity dispersions of $\sim$30-50~km~s$^{-1}$. The FE filament shows lower velocities and narrower dispersion ($\sim-$60~km~s$^{-1}$, $\sigma\sim$30~km~s$^{-1}$). 

            \item We computed VSFs across localized subregions of the filament network. Most VSFs show steep slopes (> $\sim$2/3) and flattening at small scales of a few hundred parsecs. These trends, consistent across the central and SE filaments, point to a common turbulence driver -- most likely SNe Ia and, if confirmed, observed here for the first time -- rather than the cascade of turbulence and/or shaped by dissipation processes or magnetic effects. In contrast, the FE filament exhibits slightly lower amplitude, suggesting weaker active driving. The proximity of M87 allows us to resolve these scales, but higher-resolution observation (e.g., MUSE NFM) are needed to probe smaller scales and to assess the robustness of these findings. In some regions, elevated VSF amplitudes or slopes are observed, likely due to LOS superposition. We also emphasize that the finite size of the analysis apertures can influence the measured flattening. Such limitations must be considered when interpreting the observed turbulence cascade.

            \item In the SE filament, CO(2–1) emission detected with ALMA coincides with the ionized gas, and both phases show comparable VSFs on small scales. This agreement is consistent with models where colder gas inherits warm-gas motions, and may reflect direct dynamical coupling between the phases. Differences at larger scales and broader ionized-gas velocity dispersions may reflect geometry  and differing volume filling factors.

            \item Emission-line diagnostics, including line ratios and WHAN diagrams, indicate that ionization in the filaments is dominated by non-stellar sources. The elevated [\ion{N}{ii}]/H$\alpha$ and [\ion{S}{ii}]/H$\alpha$ ratios, especially near the radio lobes, are consistent with AGN-related processes such as slow shocks, and may also arise from cosmic ray heating, turbulent mixing layers or X-ray emitting gas from the cooling ICM that is triggered by AGN feedback. The FE filament shows lower excitation but exceeds levels expected from old stellar populations, consistent with fossil AGN activity or low-energy processes.

            \item We confirm a strong spatial correlation between H$\alpha$ and soft X-ray surface brightness in the main filaments, as seen in other cool core clusters, implying close coupling between warm and hot phases. This behavior may result from thermal instabilities in the ICM or more generally from the interaction between cold gas and the surrounding hot atmosphere. However, this correlation weakens in the isolated FE filament and in an outer clump in Centaurus, both with lower X-ray/H$\alpha$ ratios ($\la$ 1), possibly representing to gas uplifted during earlier AGN activity that has since cooled more efficiently, evolved differently, or become less coupled to the ambient hot phase. However, we note that this discrepancy could also reflect pGMCA limitations.
        \end{enumerate}

    Overall, this study demonstrates the value of combining spatially resolved spectroscopy with multi-phase to investigate AGN-driven filamentary structures in the core of M87. By targeting both central and more distant regions (e.g., FE filament), we highlight how local conditions and feedback history shape the dynamics, excitation, and coupling of the multiphase gas.

\section*{Acknowledgements}
    We thank GTC staff for their flexibility (allowing visitor mode observations), help and  support for data acquisition, quality assessment and reduction (D. Reverte Payá), as well as co-observer R. Varas González for his support during the GTC observing run.
    The authors acknowledges that this work is based on observations obtained with SITELLE, a joint project of Université Laval, ABB, Université de Montréal, and the Canada-France-Hawaii Telescope (CFHT), which is operated by the National Research Council (NRC) of Canada, the Institut National des Sciences de l'Univers of the Centre National de la Recherche Scientifique (CNRS) of France, and the University of Hawaii. The observations were carried out with care and respect from the summit of Maunakea, a site of great cultural and historical significance.
    This paper makes use of the following ALMA data: ADS/JAO.ALMA\#2013.1.00862.S. ALMA is a partnership of ESO (representing its member states), NSF (USA) and NINS (Japan), together with NRC (Canada), NSTC and ASIAA (Taiwan), and KASI (Republic of Korea), in cooperation with the Republic of Chile. The Joint ALMA Observatory is operated by ESO, AUI/NRAO and NAOJ.
    C.P. acknowledges support from the FRQNT master's training scholarship (https://doi.org/10.69777/346450).
    M.-L.G.-M. acknowledges financial support from the grant CEX2021-001131-S funded by MCIU/AEI/ 10.13039/501100011033, from the coordination of the participation in SKA-SPAIN, funded by the Ministry of Science, Innovation and Universities (MCIU), as well as NSERC via the Discovery grant program and the Canada Research Chair program.
    V.O. acknowledges support from the DYCIT ESO-Chile Comite Mixto PS 1757, Fondecyt Regular 1251702.
    Y.L. acknowledges support from NASA grant 80NSSC22K0668, Chandra X-ray Observatory grant TM3-24005X, NSF grants AST-2514692, AST-2510198, and CAREER award AST-2516092.
    L.H.M. acknowledges finantial support by the grant PID2021-124665NB-I00 funded by the Spanish Ministry of Science and Innovation and the State Agency of Research MCIN/AEI/10.13039/501100011033 PID2021-124665NB-I00 and ERDF A way of making Europe.
    S.C. acknowledges the I+D+i project PID2022-140871NB-C21, financed by MICIU/AEI/10.13039/501100011033/ and "FEDER/UE".

\section*{Data Availability}
    The data underlying this article will be shared on reasonable request to the authors.

\bibliographystyle{mnras}
\bibliography{references}


\appendix
    \section{Stellar Continuum Spectra}
    \label{appendix:stellar_continuum_spectra} 
\begin{figure*}
                \centering
                \includegraphics[width=\textwidth]{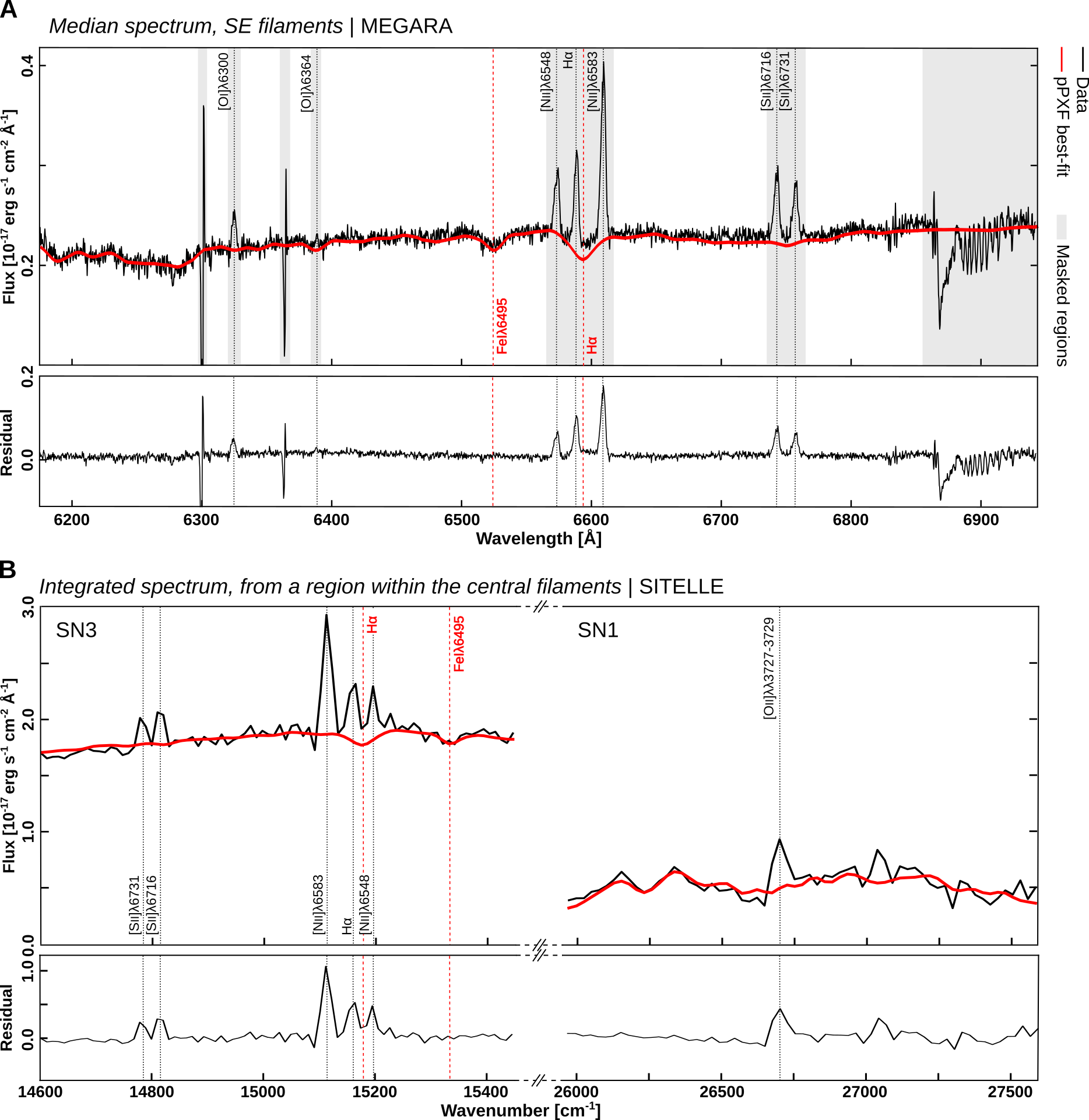}
                \caption{\small (A) Median spectrum of all spaxels in the MEGARA SE pointing, excluding edge regions affected by instrumental artifacts. Shaded areas indicate masked regions omitted from the pPXF fit (see main text for details). (B) Integrated spectrum from a region within the central filament, extracted from SITELLE data at ($\alpha$: 187.7075991: $\delta$: 12.3938644)$\degr$ with a circular aperture of radius $\sim1.4$$\arcsec$. The spectrum show data from the SN3 and SN1 filters, separated by a broken wavenumber axis but fitted simultaneously. In both panels, the best-fit stellar continuum model from pPXF is shown in red, with residuals plotted below. Main emission lines are marked by vertical black dotted lines, while major stellar absorption features are indicated in red dashed lines. In panel A, the [\ion{O}{I}]$\lambda$6300 line is visible, but too faint to be measured reliably on a spaxel-by-spaxel basis (SNR < 3); its line width appears consistent with the other emission features.}
                 \label{fig:pPXF-Spectra}
            \end{figure*}

        Figure~\ref{fig:pPXF-Spectra} presents representative spectra and their corresponding stellar continuum fits used for continuum subtraction in the MEGARA and SITELLE data.
        
    \section{SITELLE Velocity Dispersion MAP}
    \label{appendix:sitelle_dispersion}
\begin{figure}
                \centering
                \includegraphics[width=\columnwidth]{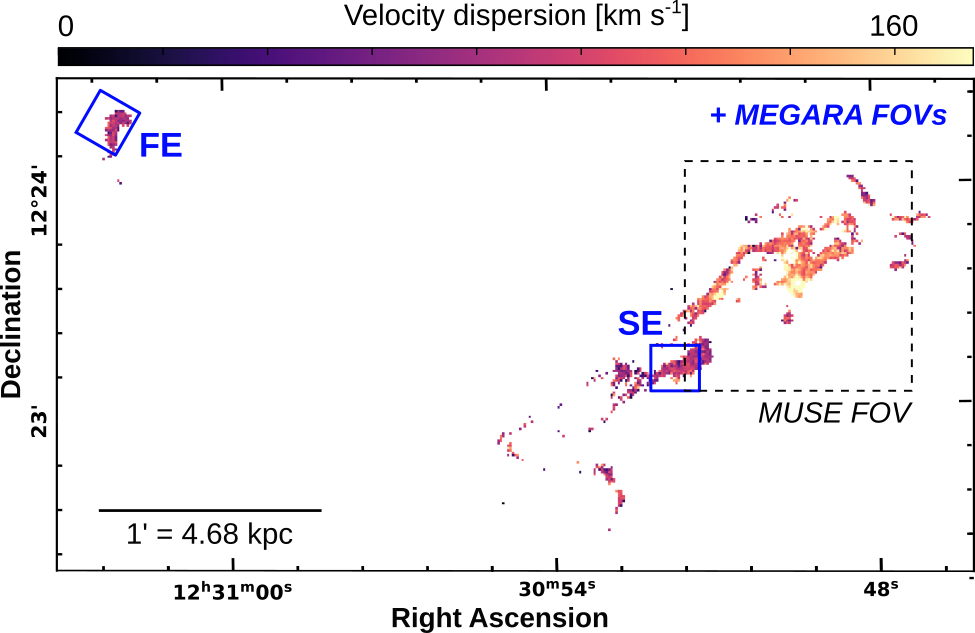}
                \caption{SITELLE velocity dispersion map showing spaxels with SNR$_{\text{H}\alpha} > 3$ and velocity uncertainties $\leq 30$~km~s$^{-1}$ (2~$\times~2$ binning). The MEGARA FOVs targeting the SE and FE filaments are outlined in blue, and the FOV of the existing MUSE WFM observations is shown in a black dotted contour.}
                \label{fig:VelocityDispersion_SITELLE}
            \end{figure}

        Figure~\ref{fig:VelocityDispersion_SITELLE} presents the velocity dispersion map derived from the SITELLE data (SN3 filter). We observe elevated values in the very center, reaching up to nearly 200~km~s$^{-1}$, followed by intermediate dispersions of approximately 125~km~s$^{-1}$ within the filaments covered by the MUSE WFM pointing. In the outer regions, including the FE filament, the dispersion drops to lower values of about $\sim$80~km~s$^{-1}$. However, we emphasize that these measurements are significantly affected by the limited spectral resolution of SITELLE, and should be interpreted with caution.
        
    \section{Impact of Filament Segmentation on VSFs}
    \label{appendix:vsf_subregions}
        To assess how the segmentation of filamentary substructures impacts the interpretation of turbulence, we performed a comparison of VSFs computed with and without applying the spatial segmentation described in Section~\ref{sec:multiple_vsfs}. The goal is to test whether separating filaments leads to significant differences in the VSF parameters. Using the methodology described in Section~\ref{sec:multiple_vsfs}, we computed localized VSFs across the MUSE velocity map with the same sampling (central positions spaced by $\sim 100$ pc and aperture radius of $\sim0.39$ kpc), but without applying any subregion boundaries. For each central region, we then compared the resulting VSF to its counterpart obtained from the segmented approach by measuring the relative differences in three key parameters: the slope (power-law index), the flattening scale and the amplitude at flattening. 
    
\begin{figure}
                \centering
                \includegraphics[width=\columnwidth]{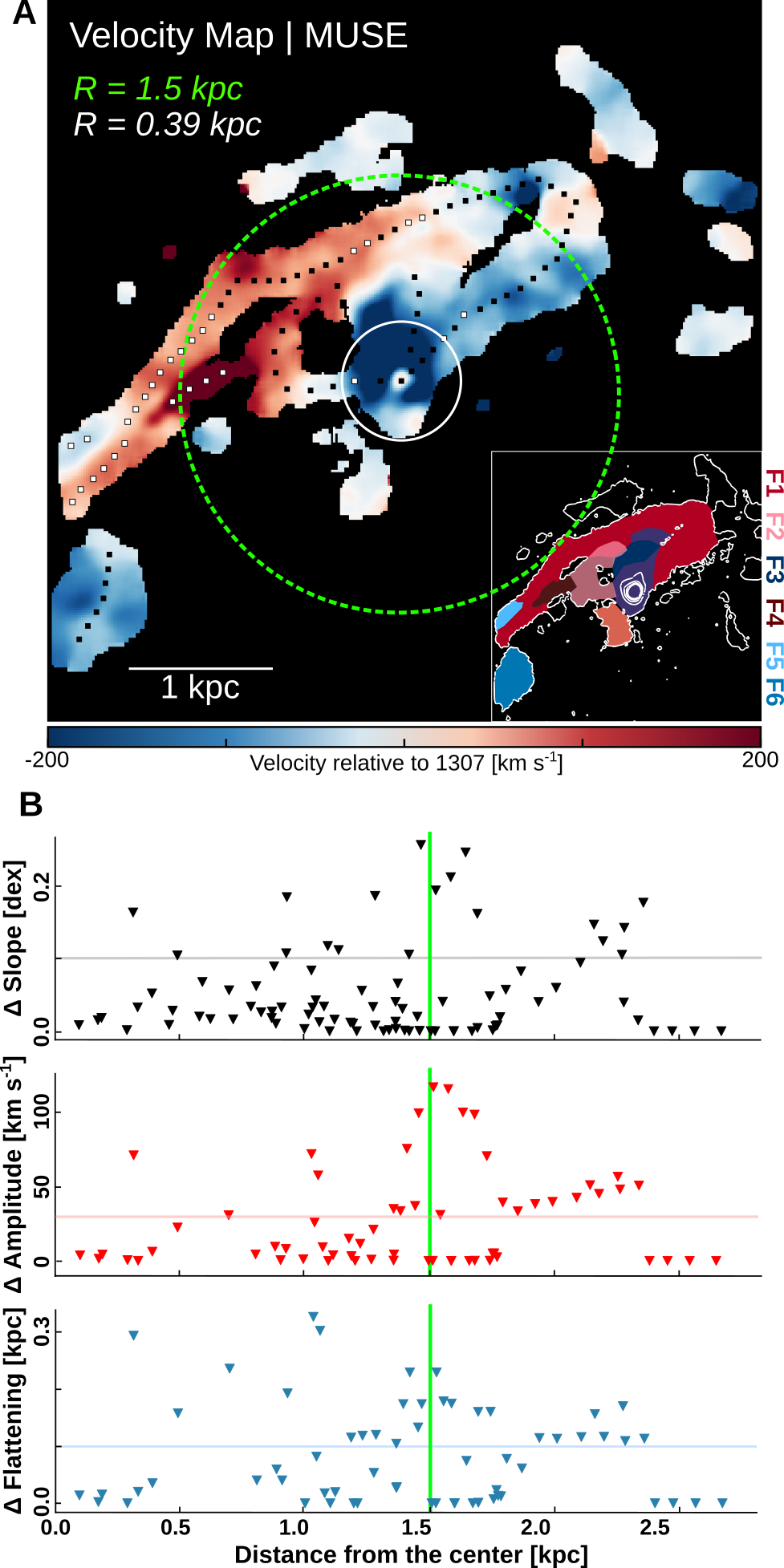}
                \caption{\small (A) MUSE velocity map of the central filaments. Each square marks the central position of the localized VSF, computed within a circular aperture of radius 0.39 kpc, with and without applying the filament segmentation. White squares indicate locations where at least one VSF parameter exceeds the defined threshold (0.1 dex for the slope, 30 km s$^{-1}$ for the amplitude and 0.1 kpc for the flattening scale). (B) Variation of the VSF parameters as a function of projected distance from the galaxy center. Horizontal lines mark the corresponding thresholds for each parameter. The vertical green line indicates the projected extent of the optical jet ($\sim 1.5$ kpc).}
                \label{fig:VSFs_MUSE_subregions}
            \end{figure}

        Figure~\ref{fig:VSFs_MUSE_subregions} shows the variation in VSF parameter differences -- slope, flattening scale, and amplitude -- as a function of projected distance from the galaxy center. The results indicate that the impact of segmentation on VSF measurements varies across the field. Differences in slope remain below 0.1 dex in approximately 80\% of the regions, while amplitude differences exceed 30 km s$^{-1}$ in 28\% of the cases. Flattening scale differences greater than 0.1 kpc are observed in about 30\% of regions. Importantly, the most significant deviations in all three parameters tend to occur in the same areas -- particularly in subregions F1, F4, and F5 -- where distinct kinematic components are known to overlap. This spatial coincidence suggests that the segmentation effectively reduces contamination from unrelated structures, leading to more reliable measurements of turbulence properties.
    
    \section{Impact of Spaxel Size on VSFs}
    \label{appendix:vsf_spaxelsize}
            \begin{figure*}
                \centering
                \includegraphics[width=\textwidth]{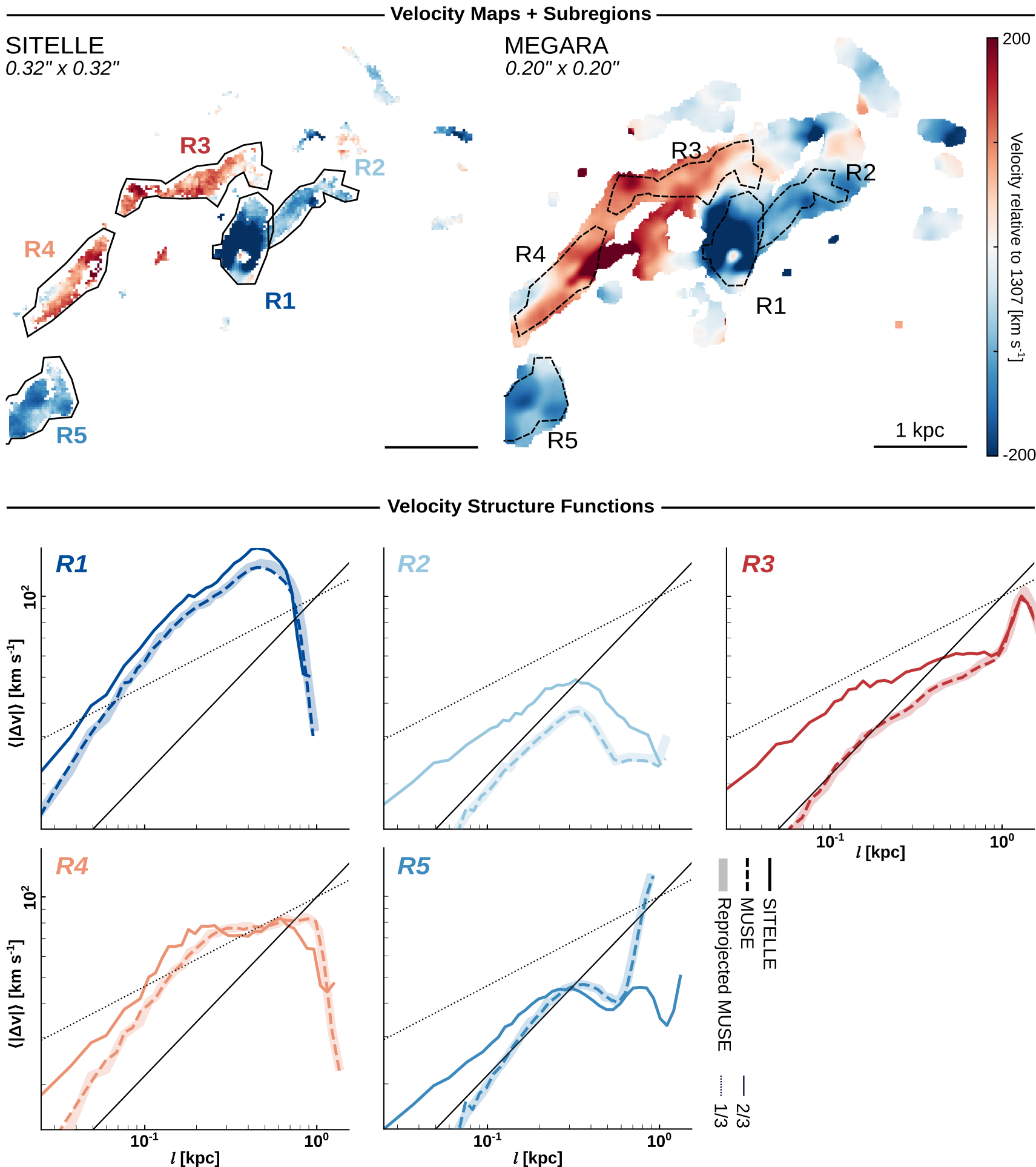}
                \caption{\small Top panels: Velocity maps of the central filaments obtained with SITELLE (left) and MUSE (right), limited to MUSE FOV. The subregions used for VSF analysis (R1 to R5) are outlined in both maps. Bottom panels: VSFs computed for each subregion using SITELLE (solid lines) and MUSE (dashed lines). Reference slopes of 1/3 (Kolmogorov turbulence) and 2/3 are shown for comparison.}
                 \label{fig:VSFs_Resolution_SITELLE-vs-MUSE}
            \end{figure*} 
        
        To asses whether differences in spaxel size between instruments impact the measurement of VSFs, we first compared SITELLE and MUSE data over matched subregions in central filaments of M87. SITELLE data have an original spaxel size of 0.32$\arcsec$~$\times$~0.32$\arcsec$ (unbinned), while MUSE offers a finer spaxel size of 0.20$\arcsec$~$\times$~0.20$\arcsec$. The SITELLE unbinned map was produced following the same procedure described in Section~\ref{sec:obs_sitelle}, but without spatial binning. To enable a direct comparison at the same spaxel size, we also resampled the MUSE velocity maps to the SITELLE grid using the \texttt{reproject} function from the \texttt{astropy} package. We manually defined five subregions (R1 to R5) in the SITELLE velocity maps, selecting areas with well-defined, continuous filamentary structures and sufficient pixel counts. The same spatial masks were then applied to the MUSE data. Owing to its finer spaxel size and higher SNR across the filaments, MUSE contains a larger number of spaxels per region. For each subregion, we computed the first-order VSF as described in Section~\ref{sec:vsf}. The results are shown in Figure~\ref{fig:VSFs_Resolution_SITELLE-vs-MUSE}. Overall, the VSF slopes are broadly consistent between SITELLE and MUSE at intermediate and large scales. However, at small separations -- particularly below the seeing limit of SITELLE ($\sim$1$\arcsec$) -- SITELLE VSFs tend to diverge more strongly. Most regions show a consistent flattening scale between instruments, with some displaying a second flattening at larger separations where the number of pairs becomes insufficient. Across all subregions, SITELLE exhibits systematically higher VSF amplitudes. The resampled MUSE data produce nearly identical VSFs to the original map, confirming that this amplitude offset is not due to spaxel size differences but likely stems from the noisier nature of the SITELLE velocity map, which shows enhanced pixel-to-pixel variations.
    
\begin{figure*}
                \centering
                \includegraphics[width=\textwidth]{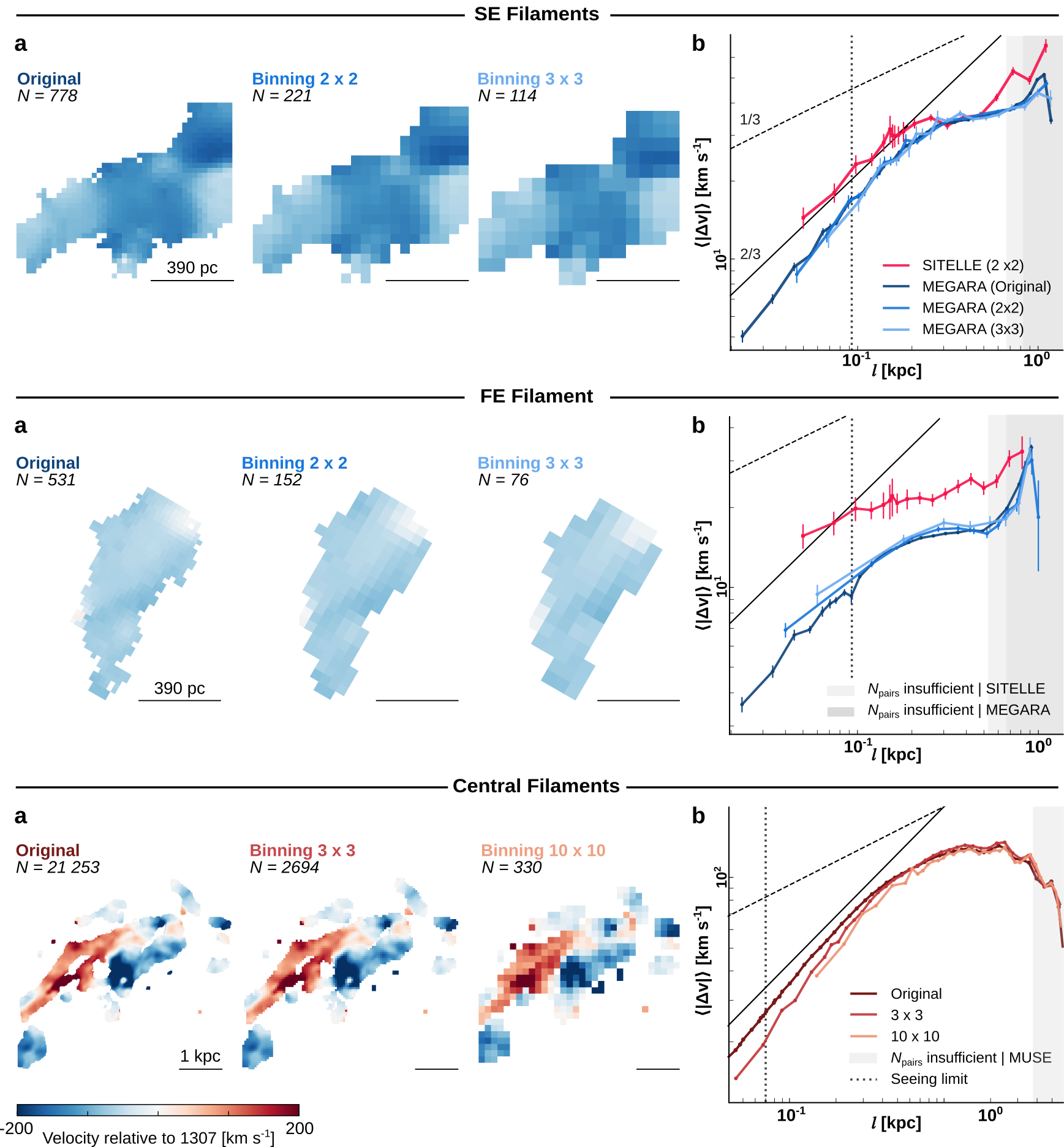}
                \caption{\small (a) Velocity maps of the original and binned datasets for the SE filaments (top panel) and FE filaments (middle panel) with MEGARA, and central filaments (bottom panel) with MUSE. The number of spaxels $N$ in each map is indicated in the upper-left corner. Color scale is consistent across maps. (b)~Corresponding VSFs derived from each map, with reference slopes of 1/3 and 2/3 shown for comparison. Shaded areas indicates bins with an insufficient number of pairs in the original maps. For the SE and FE filaments, SITELLE VSFs (2~$\times$~2) are also shown for comparison. The typical seeing limit (SITELLE: $\sim$1.2$\arcsec$, MUSE: 0.9$\arcsec$) is marked by a vertical dotted line.}
                \label{fig:VSFs_Binning_SIT-MEG_MUSE}
            \end{figure*}

        In a second test, we assessed the impact of spatial binning on the shape of the VSFs. For MEGARA, the binning was applied directly to the data cube by averaging the spectra within 2~$\times$~2 and 3~$\times$~3 spaxels, followed by re-fitting the binned spectra with a single-component model, as described in Section~\ref{sec:emission_line_fitting}, to generate new velocity maps. For MUSE, where only the velocity map was available, binning was performed by averaging the velocity values within 3~$\times$~3 and 10~$\times$~10 pixel regions. VSFs were then recomputed for each binned velocity map using the same methodology as described previously. Figure~\ref{fig:VSFs_Binning_SIT-MEG_MUSE} presents the original and binned velocity maps, along with their corresponding VSFs. For the SE and FE filaments, the SITELLE VSFs (binning of 2~$\times$~2) are also shown for reference. We find that binning has minimal effect on the overall shape of the VSFs. The flattening scale and its amplitude remain stable across all binning levels, including the extreme case of 10~$\times$~10 binning in MUSE (effective spaxel size $\sim$2$\arcsec$). The primary effect is a steepening of the slope at small scales, likely due to enhanced contrast between adjacent spaxels in the binned maps. Unlike Gaussian seeing, which smooths gradients, binning tends to preserve or amplify sharp transitions in the velocity field. These findings reinforce our conclusion that binning has moderate influence on VSF shape and amplitude. Once again, SITELLE systematically shows higher amplitudes than MUSE and MEGARA -- even within comparable FOVs -- further suggesting that the underlying data quality, rather than spatial sampling, drives this discrepancy.
            
    \section{Impact of the window size on VSFs}
    \label{appendix:vsf_fov}
            \begin{figure*}
                \centering
                \includegraphics[width=\textwidth]{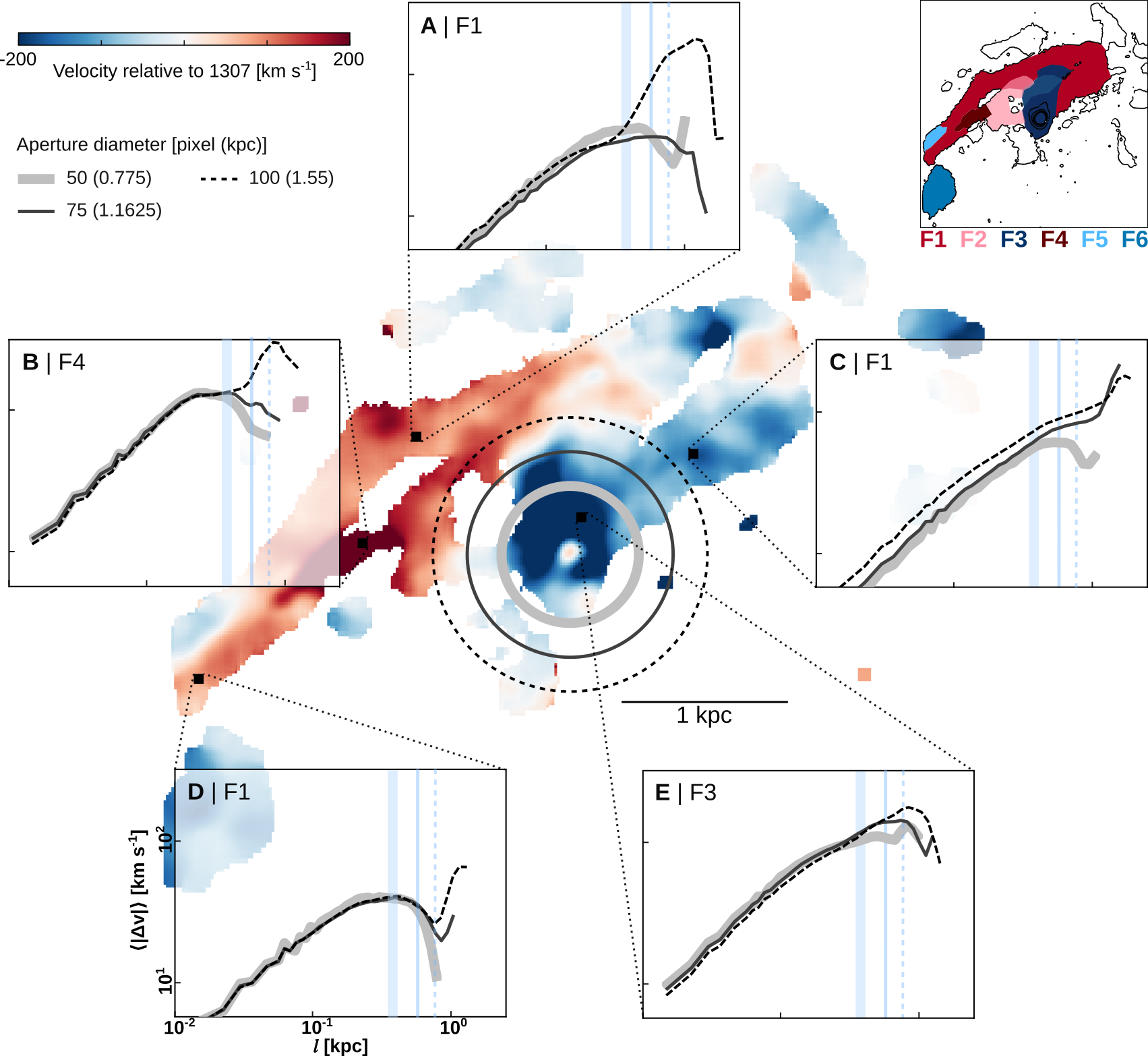}
                \caption{\small The background shows the MUSE velocity map, with its scale bar located at the bottom right of the three circular apertures. Insets A to E display the VSFs computed within circular apertures of varying diameters -- 50 (light grey), 75 (medium grey) and 100 spaxels (black dashed line) -- centered on the black squares in a subregion (labelled F1 to F5). These subregions are defined in the reference map at the top right. The relative size of each aperture with respect to to the filament is illustrated by circles on the galaxy nucleus. Tick labels are displayed only in inset D; all insets share the same axis scaling. Vertical blue lines mark half the size of each aperture’s spatial window, with line styles matching those of the corresponding circular apertures.}
                \label{fig:VSFs_FOVSize_MUSE}
            \end{figure*}
    
        To assess whether window size affects the measurement of VSFs, we performed a test using the full MUSE velocity map of the central region of M87. This approach is based on the subregions defined in Section~\ref{sec:multiple_vsfs}. We defined a series of circular FOVs centered at representative locations, with diameter of 50, 75 and 100 spaxels (corresponding to 0.775~kpc, 1.1625~kpc and 1.55~kpc) and computed the first-order VSF for each in their subregion using the method described in Section~\ref{sec:vsf}. The results are shown in Figure~\ref{fig:VSFs_FOVSize_MUSE}.
    
        We find that the overall shape of the VSFs does not show significant changes on smaller scales across different aperture sizes. Both the slope and the location of the first flattening are generally stable, indicating that small- and intermediate-scale turbulence is not strongly sensitive to our selected window size. However, as the aperture increases, the VSF extends to larger separations, and a second rise and flattening becomes more pronounced -- particularly in cases such as insets A/B, where the first flattening gradually becomes less visible as large-scale motions begin to dominate. In other cases, such as inset C, the first flattening becomes less distinct and is replaced by a smoother slope with increasing FOV.
    
        This test confirms that, within the range of aperture sizes considered here, the measurement of VSFs is relatively robust up to scales of $\la$2~kpc. Nevertheless, larger FOVs increasingly may sample different dynamical regimes, shifting the dominant contribution from localized turbulence to more global kinematic structures. We also note that the window size can introduce artificial features in the VSF, typically at scales around half its length, although this effect is not always significantly apparent here. As such, care must be taken when interpreting the behaviour of the VSF or when comparing results across datasets with substantially different spatial coverage.


\bsp	
\label{lastpage}
\end{document}